\documentclass[11pt]{article}
\usepackage{amsfonts}
\usepackage{amssymb}
\usepackage{amsmath}
\usepackage{amscd}
\usepackage[all]{xy}
\begin{document}

\thispagestyle{empty}

\begin{center}
CENTRE DE PHYSIQUE TH\'EORIQUE \footnote{\, Unit\'e Mixe de
Recherche (UMR) 6207 du CNRS et des Universit\'es Aix-Marseille 1 et 2 
\\ \indent \quad \, Sud Toulon-Var, Laboratoire affili\'e \`a la
FRUMAM (FR 2291)} \\ CNRS--Luminy, Case 907\\ 13288 Marseille Cedex 9\\
FRANCE
\end{center}

\vspace{1.5cm}

\begin{center}
{\Large\textbf{Jets, frames, and their Cartan geometry}}
\end{center}

\vspace{1.5cm}

\begin{center}
{\large Micha\"el Grasseau
\footnote{\, grasseau@cpt.univ-mrs.fr} }

\vspace{1.5cm}

{\large\textbf{Abstract}}
\end{center}

Jet frames, that is a generalisation of ordinary frames on a
manifold, are described in a language similar to that of gauge theory. 
This is
achieved by constructing the Cartan geometry of a manifold with respect to the
diffeomorphism symmetry. This point of vue allows to give new insights and
interpretations in the theory of jet frames, in particular by making an
interpolation between ordinary gauge theory concepts and pure jet ones. 

\vspace{1.5cm}

\vskip 1truecm

\vskip 1truecm

\noindent March 2006

\newpage 

\sloppy
\def\got#1{{\mathfrak{#1}}}
\def\bad#1{{\overline{\rm{#1}}}}
\def\rec#1{{\rm{#1}}}

{\bf Introduction}
\\ \\
A description of jet theory, and more precisely that of jet frames,
described e.g. in \cite{koba} or \cite{pom}, is proposed on the basis of 
Cartan type geometry :  
the geometry associated to a differentiable manifold
$M$ formally represented as the homogeneous space 
$$ M \simeq \rec{Diff} (M) / \rec{Diff}_x (M) $$
where $\rec{Diff}_x(M)$ are the diffeomorphisms that don't move a point
$x \in M$, is constructed. 
\\ \\
The interest of such a construction is that it realises a intermediate between
the pure jet language \cite{pom} and the pure gauge theory language (principal
fiber bundles). This gives an alternative description, in global terms, of the
differential sequences given in \cite{pom}, a gravity interpretation of the
objects introduced, all being synthetised in some field theory of frames.
\\ \\
The first section, needed for both technical and notational purposes, is a short review
and reformulation of the algebraic machinery exposed in \cite{koba}, and
alternatively in \cite{cap} and \cite{baston} in a closely related context. 
\\ \\
The second section begins by recalling what are the jet frames of \cite{koba},
or, as we shall see of \cite{pom}. We then describe an alternative viewpoint
on the subject, based on a procedure of prolongation similar to that of
\cite{koba} or \cite{cap}, but here adapted to the infinite dimensional
geometry of $\rec{Diff}(M)$. It allows to construct the so-called linear
frames, of arbitrary order, the first order frames being the usual ones. See
\cite{newman} for an example of the use of Cartan connection, i.e. the dual
version of 2-frames and 3-frames there, in gravity. 
\\ \\
The third section presents a field theory like treatment of the objects thus
constructed. It is shown how to recover, in a simplified manner, the
differential operators and sequences of \cite{pom}, and a concrete
description is given, in terms of symmetry, and deformations. 

\newpage
\tableofcontents 

\section{Algebraic preliminaries}

Two functions $\phi, \phi' : \mathbb{R}^n \to \mathbb{R}^m$ are said equivalent
to order $k$ at $x \in \mathbb{R}^n$ if they have the same derivatives at $x$ up to
order $k$. The equivalence class is called a $k$-jet, and denoted $j^k_x (\phi)$.    

\subsection{Formal vector fields, Jet groups}

$\bullet$ 
On $\mathbb{R}^n$ with coordinates $x^a$, $a=1,\cdots,n$, the formal vector
fields are the ($\partial_a = \frac{\partial}{\partial x^a}$ and sum on
repeted index)
$$
       X = \sum_{k \geq -1} X_{k} \,\,\, {\rm with} \,\,\, 
       X_{k} = \frac{1}{k!} X^a {_{b_1 \cdots b_{k+1}}} 
       x^{b_1} \cdots x^{b_{k+1}} \partial _{a} 
$$       
equiped with minus the ordinary Lie bracket of vector fields (the minus is taken
by analogy with a group acting on one of its homogeneous space, see
\cite{koba}). This defines a graded Lie algebra 
$$
   \got{gl}_{\infty} = \bigoplus_{k \geq -1} \got{gl}_k \,\,\, 
   {\rm with} \,\,\, 
   \left[ \got{gl}_k , \got{gl}_{k'} \right] \subset \got{gl}_{k+k'} 
$$
where $\got{gl}_k$ is the space of $X_k$'s. The $k$'s are "spins" with respect
to the dilatation operator 
$$  [X_k , D] = k X_k , \; D = x^a \partial_a $$
$\bullet$ 
The jet group $GL^k$ of order $k$ is the space of $(k+1)$-jets of (orientation
preserving) local 
diffeomorphisms $g$ of $\mathbb{R}^n$ such that $g(0)=0$. Denoting by
$g^k = j^{k+1}_0 (g)$ its elements, the group law is (formal successive
derivations) 
$$ g^k g'^k = j^{k+1}_0 (g \circ g') $$
By restrictions on the order of jets, we obtain projections $GL^k \to GL^{k-1}$
whose kernel $GL_k$ is normal and abelian in $GL^k$, and we have 
$$ GL^k / GL_k \simeq GL^{k-1} , \; GL^k \simeq GL^{k-1} \ltimes GL_k $$
Recursively, the projections $GL^k \to GL^{k-1} \to \cdots \to GL^0 = GL_0 $
induce the decomposition (factorisation of jets) 
$$ 
      GL^k = GL^{k-1} \ltimes GL_{k} = ( GL^{k-2} \ltimes GL_{k-1} ) 
      \ltimes GL_{k} = \cdots 
$$
and we shall denote this $GL^k = GL_0 \ltimes GL_1 \ltimes \cdots \ltimes GL_k$
, in correspondance with the decomposition $g^k = g_0 g_1 \cdots g_k$. 
\\ 
Alternatively, letting $H^k$ be the $\infty$-jets such that $j^{k+1}_0 (g) =
j^{k+1}_0 (\rec{id})$, we obtain a normal subgroup of $GL^{\infty}$ which
identifies $GL^k \simeq GL^{\infty} / H^k$. So, infinitesimally, we obtain the
Lie algebra isomorphisms 
$$
    {\rm Lie} H_{k} =  \bigoplus_{l \geq k+1} \got{gl}_l , \;   
    {\rm Lie} G_k  =  \bigoplus_{l \geq 0} \got{gl}_l / \bigoplus_{l \geq k+1} \got{gl}_l \simeq
                    \bigoplus_{l \geq 0}^{k} \got{gl}_l
$$
So, the product in $GL^k$ is the truncation to $(k+1)$-jets of the product in
$GL^{\infty}$. 

\subsection{The jet action $\bad{Ad}$}
\label{sectionthejetaction}

For $X \in \got{gl}_{-1} \oplus \cdots \oplus \got{gl}_k$ written 
$ X = \left. \frac{d}{dt} \right. _{|t=0} j^{k+1}_0 (\phi_t) $ where $\phi_t :
\mathbb{R}^n \to \mathbb{R}^n$ for each $t$ on the path $t
\to \phi_t$, $\phi_0 = \rec{id}$, and $g^{k+1} = j^{k+2}_0 (g)$, $g(0) = 0 $,
define : 
\begin{equation}
    \bad{Ad}(g^{k+1}) X = \left. \frac{d}{dt} \right._{|t=0} j^{k+1}_0 (g
    \circ \phi_t \circ g^{-1}) 
\end{equation}
This is well defined since the result only depends on the $(k+2)$-jet of $g$.
This is an action of $GL^{k+1}$ on $\got{gl}_{-1} \oplus \cdots \oplus
\got{gl}_k$. 
In particular $\bad{Ad} (g_{k+1})$, $ g_{k+1} \in GL_{k+1} $ is an isomorphism
of degree $k$ of $\got{gl}_{-1} \oplus \cdots \oplus \got{gl}_k$ : 
\begin{equation}
   \bad{Ad} (g_{k+1}) \left( X_{-1} \oplus \cdots \oplus X_{k} \right) 
   = X_{-1} \oplus \cdots \oplus X_{k-1} \oplus X_{k} + \alpha_k (X_{-1}) 
   \label{degreek}
\end{equation}
where $\alpha_k \in \got{gl}_{k+1} \subset \got{gl}_{k} \otimes \got{gl}_{-1}^*$
thanks to $GL_{k+1} \simeq \got{gl}_{k+1}$, $k \geq 0$. 
We denote by $GL_{k,1}$ the group of degree $k$ isomorphisms of 
$\got{gl}_{-1} \oplus \cdots \oplus \got{gl}_k$, then $GL_{k,1} \simeq
\got{gl}_k \otimes \got{gl}_{-1}^*$, its action being given by the same formula 
(\ref{degreek}). Finally, we obtain in this way an action of 
$GL^k \ltimes GL_{k,1}$ on $\got{gl}_{-1} \oplus \cdots \oplus
\got{gl}_k$, which extends $\bad{Ad}$, and still denoted $\bad{Ad}$. 

\subsection{Spencer cohomology}

Spencer cohomology \cite{baston} is the cohomology of the abelian Lie algebra of translations
$\got{gl}_{-1}$ with values in $\got{gl}_{\infty}$, so Spencer cochains are
$ \got{gl}_{\infty} \otimes \Lambda^* \got{gl}_{-1}^*$. This space
decomposes into a direct sum of the $\got{gl}_{k,l} = \got{gl}_k \otimes
\Lambda^l \got{gl}_{-1}^*$. For a cochain $\alpha$ of form degree $l$, the coboundary
operator is 
\begin{equation}
    \partial \alpha = 
    \sum_{i=0}^{l}{ (-1)^i [X_i, \alpha (X_0 , \cdots , \hat X_i , \cdots , X_l)
    ]} , \,\,\, X_i \in \got{gl}_{-1} , \,\,\, \partial^2 = 0     
\end{equation}
where $\hat{}$ here denotes omission. 
In particular $\got{gl}_{k+1}$ appears as the kernel of 
$ \got{gl}_{k,-1} \stackrel{\partial}{\longrightarrow}
\got{gl}_{k-1,2} $. More generally, Spencer
$\partial$-cohomology is trivial \cite{pom}, and so the particular sequences (for each $k$)
\begin{equation}
     \xymatrix{ 
     0 \ar[r] & 
     \got{gl}_{k+1} \ar[r] & 
     \got{gl}_{k,1} \ar[r]^-\partial & 
     \got{gl}_{k-1,2} \ar[r]^-\partial & 
     \cdots \ar[r]^-\partial & 
     \got{gl}_{k-n+1,n} \ar[r] & 0 
     }
\end{equation}
are exacts. 
\\
The $\bad{Ad}$ action of $GL^{k+1}$ on $\got{gl}_{-1} \oplus \cdots \got{gl}_k$
induces an action on Spencer cochains, that we still denote $\bad{Ad}$, and
given by, for $\alpha = \alpha_{-1} \oplus \cdots \oplus \alpha_k \in
\got{gl}_{-1,l} \oplus \cdots \oplus \got{gl}_{k,l}$ :
$$
    \bad{Ad} (g) \alpha = \bad{Ad} (g) \circ \alpha \circ \bad{Ad} ({g_0}^{-1})
    , \; g \in GL^{k+1} , \; g = g_0 . g_1 . \cdots . g_{k+1} 
$$

\subsection{Notations}

For $G$ a Lie group, a $G$-principal bundle $P$ above the base space $M$ will be
denoted by
$$ 
   \xymatrix{ G \ar[r] & P \ar[r] & M } 
$$
We shall think of this as a non linear version of a short exact sequence. For $g
\in G$, the right action on $p \in P$ is denoted $R_g (p) = p.g$, and the
vertical vector field on $P$ induced by $X \in \rec{Lie} G$ is denoted $\hat X$.
\\ 
The associated bundle $E$ defined by a left action $\rho$ of $G$ on the space $V$ will
be denoted 
$$ 
      E = P \times_{\rho}  V
$$
and its space of sections $\Gamma(E)$. 
The space of $l$-forms on $M$ with values in the bundle $E$ is denoted
$\Omega^l (M,E)$, and the space of tensorial forms on $P$ with values in $V$ is denoted
$\Omega^l_G (P,V)$. These two spaces are isomorphic.       
       
\section{Geometry of frames}

Fix now an $n$-dimensional differentiable (and orientable) manifold $M$.   

\subsection{Jet frames}

A $(k+1)$-jet frame above $x \in M$ is the $(k+1)$-jet at $0$ of a (orientation
preserving) local diffeomorphism
$\phi : \mathbb{R}^n \to M$ such that $\phi(0) = x$. We shall denote this $e^{k} =
j^{k+1}_0 (\phi)$, and $M^k$ the space of $e^k$'s. The projection 
$$ 
   \pi_{k,-1} : M^k \to M , \; e^k \mapsto x
$$
where $e^k = j^{k+1}_0(\phi)$, $x= j^0_0(\phi)= \phi(0)$, and right action
$$
    M^k \times GL^k \to M^k , \; ( e^k , g^{k} ) 
    \mapsto R_{g^k} (e^k) = e^k. g^k = j^{k+1}_0 (\phi \circ g) 
$$
where $e^k=j^{k+1}_0(\phi)$, $g^{k}=j^{k+1}(g)$ with $g(0)=0$, 
turns $M^k$ into $GL^k$-principal bundle above $M$ : 
\begin{equation}
   \xymatrix{ 
   GL^k \ar[r] & M^k \ar[r] & M 
   }
\end{equation}
More generally, for $k' < k$, the projection 
$$ 
   \pi_{k,k'} : M^k \to M^{k'} , \; e^{k} \mapsto 
   e^{k'} 
$$
where $e^{k'}=j^{k'+1}(\phi)$, 
and right action 
$$
    M^k \times GL_{k'+1} \ltimes \cdots \ltimes GL_k \to M^k , \; 
    ( e^k , g^{k'k}  ) 
    \mapsto R_{g^{k'k}} (e^k) = e^k. g^{kk'} = j^{k+1}_0 (\phi \circ g)    
$$
where $g^{k'k} = j^{k+1}_0(g)$ with $j^{k'+1}_0(g)=j^{k'+1}_0(\rec{id})$,            
defines on $M^k$ the structure of a $GL_{k'+1} \ltimes \cdots \ltimes
GL_k$-principal bundle above $M^{k'}$ : 
\begin{equation}
   \xymatrix{ 
   GL_{k'+1} \ltimes \cdots \ltimes GL_k \ar[r] & M^k \ar[r] & M^{k'} 
   }
\end{equation}
We obtain in this way a tower of principal bundles : 
\begin{equation}
   \xymatrix{
   M^k \ar[r] & M^{k-1} \ar[r] & \cdots \ar[r] & M^0 \ar[r] & M 
   }
\end{equation}
Alternatively, since $GL_{k'+1} \ltimes \cdots \ltimes GL_k$ is a normal
subgroup of $GL^k$, we have an induced principal structure on the quotient 
$ M^k / GL_{k'+1} \ltimes \cdots \ltimes GL_k $ and this is isomorphic with
$M^{k'}$. See e.g. \cite{koba} for a coordinate description of these bundles. 
\subsection{Interpretation : Induced linear frames}
\label{sectioninterpretation}

Let $k \geq -1$. Denoting by $\mathbb{R}^{n,k}$ the $(k+1)$-jet frames bundle of $\mathbb{R}^n$,
and $O = j^{k+1}_0 (\rec{id})$, we obtain the natural isomorphy : 
$$
    T_O \mathbb{R}^{n,k} \simeq \got{gl}_{-1} \oplus \cdots \oplus \got{gl}_k  
$$
because each $X = X_{-1} \oplus \cdots \oplus X_k \in \got{gl}_{-1} \oplus
\cdots \oplus \got{gl}_k$ can be written 
$ X = \left. \frac{d}{dt} \right._{|t=0} j^{k+1}_0 (\phi_t) $.      	
\\
A $(k+2)$-jet frame $e^{k+1} = j^{k+2}_0 (\phi)$ induces a locally defined
isomorphism 
$$ 
    \overline{\phi}_{k+1} : \mathbb{R}^{n,k} \to M^{k} , \; 
         j^{k+1}_0 (f) \mapsto j^{k+1}_0 (\phi \circ f)
$$
whose derivative $\overline{\phi}_{k+1}{_*}$ at $O$ only depends on 
$j^{k+2}_0(\phi) = e^{k+1}$. 
So, to each $e^{k+1}$, we can associate the isomorphism 
$$ 
    e_{k+1} =\overline{\phi}_{k+1}{_*}{_{|O}} 
    : \got{gl}_{-1} \oplus \cdots \oplus
     \got{gl}_{k} \to T_{e^{k}} M^k 
$$
We call the $e_{k+1}$'s linear frames (of order $k+2$). 
The definition of projections $\pi_{k,k-1}$, and (infinitesimal) right action of
$ M^{k} \to M $, show successively that $e_{k+1}$ satisfies : 
\begin{eqnarray*}
     &(i)& \pi_{k,k-1}{_*} e_{k+1} \left( X_{-1} \oplus \cdots \oplus X_k
     \right) = e_{k} \left( X_{-1} \oplus \cdots \oplus X_{k-1} \right) 
     \\
     &(ii)& e_{k+1} \left( X_0 \oplus \cdots \oplus X_k \right) = 
     \hat X_{0} \oplus \cdots \oplus \hat X_k  
\end{eqnarray*}
The properties $(i)$ and $(ii)$ above means respectively the right and left
squares in the following diagram commute : 
\begin{equation*}
    \xymatrix{
    \got{gl}_0 \oplus \cdots \got{gl}_k \ar[r] \ar[d]^{\hat{}} & 
    \got{gl}_{-1} \oplus \cdots \oplus \got{gl}_k \ar[r] \ar[d]^-{e_{k+1}} &
    \got{gl}_{-1} \oplus \cdots \oplus \got{gl}_{k-1} \ar[d]^-{e_k} \\
    T_0 M^k \ar[r] & T_{e^k} M^k \ar[r] & T_{e^{k-1}} M^{k-1}
    }
\end{equation*}
where $T_0 M^k$ is the vertical tangent space of $M^k \to M$. 
Under the action of $g^{k+1} \in GL^{k+1}$, $g^{k+1} = j^{k+2}_0 (g)$,
$\phi_{k+1}$ becomes $\phi'_{k+1}$ with : 
\begin{eqnarray*}
     \phi'_{k+1} (j^{k+1}_0(f))  
     & = & 
     j^{k+1}_0 (\phi \circ g \circ f) = j^{k+1}_0 
     (\phi \circ g \circ f \circ g^{-1} \circ g ) \\
     &=& 
     \phi_{k+1} (j^{k+1}_0 (g \circ f \circ g^{-1})) . g^{k} = 
     (R_{g^k} \circ \phi_{k+1}) (j^{k+1}_0 (g \circ f \circ g^{-1})) 
\end{eqnarray*}
so, by derivation at $O$, we obtain the transformation of $e_{k+1}$ : 
\begin{equation}
    e'_{k+1} = R_{g^k}{_*} e_{k+1} \circ \bad{Ad} (g^{k+1}) 
    \label{transfjetk}
\end{equation}
         
\subsection{Frame forms and their Structure equations}

\subsubsection{Frame form}	 
		 
On $M^{k+1}$, let $u$ be a tangent vector at $e^{k+1} = j^{k+2}_0 (\phi)$,   
$$ 
   u = {\left. \frac{d}{dt} \right.}_{|t=0} j^{k+2}_{0} (\phi_t) \in T_{e^{k+1}}
   M^{k+1}
$$
where $t \to \phi_t$ a path such that $\phi_0 = \phi$. From the jet point of vue, we
define the frame form $\theta^k$ as   
\begin{equation}
    \theta^k (u) = {\left. \frac{d}{dt} \right.}_{|t=0} j^{k+1}_{0} (\phi^{-1}
    \circ \phi_t)  
    \label{def1}
\end{equation}
From the linear frame point of vue, the frame form is defined as     
\begin{equation}
    \theta^k (u) = {e_{k+1}}^{-1} \pi_{k+1,k}{_*} u 
    \label{def2}
\end{equation}
where $e_{k+1}$ is the linear frame induced by $e^{k+1}$.    
Formulas (\ref{def1}) and (\ref{def2}) agree since 
\begin{eqnarray*}
    {e_{k+1}}^{-1} \pi_{k+1,k}{_*} u 
    &=& 
    {\overline{\phi}_{k+1}}^{-1}{_*} \left. \frac{d}{dt} \right._{|t=0} 
      j^{k+1}_0 (\phi_t ) 
    = 
    \left. \frac{d}{dt} \right._{|t=0} 
      \left( {\overline{\phi}_{k+1}}^{-1} (j^{k+1}_0 (\phi_t )) \right) \\
    &=& 
    \left. \frac{d}{dt} \right._{|t=0} j^{k+1}_0 (\phi^{-1} \circ \phi_t) 
\end{eqnarray*}    
The properties of the frame form $\theta^k$ on $M^{k+1}$ are summarised in, see
\cite{koba} : 
\\ \\
{\it
The frame form $\theta = \theta^k$ on $M^{k+1}$ is a 
$\got{gl}_{-1} \oplus \cdots \oplus \got{gl}_k$ valued one-form on 
$M^{k+1}$ such that : 
\begin{eqnarray*}
    &(i)& R_{g}{^*} \theta = \bad{Ad} (g^{-1}) \theta , \; g \in GL^{k+1} \\
    &(ii) & \theta (\hat X_0 \oplus \cdots \oplus \hat X_{k+1}) = X_0
    \oplus \cdots X_{k} , \,\,\, \ker \theta = \ker \pi_{k+1,k}{_*} = T_{k+1}
    M^{k+1} \\
    &(iii)& \pi_{k+1,k}^{*} \theta^{k-1} = \theta^k \mod \got{gl}_k 
\end{eqnarray*}
}
\\ 
Properties $(i)$ and $(iii)$ follow directly from (\ref{def1}) and the definition of the 
right action and projection, and $(ii)$ is a direct consequence of (\ref{def2}) 
and the fact that $e_{k+1}$ is an isomorphism.
We will sometimes omit the superscript $k$ on $\theta^k$ when it is possible to
do so. 
The frame form decomposes as  
$\theta^k = \theta_{-1} \oplus \theta_0 \oplus \cdots \oplus \theta_k$ with
$\theta_l$ the $\got{gl}_l$ part. 
\\ \\ 
In the limit $k \to + \infty$, we can think of the frame form as the
Maurer-Cartan form on the group $\rec{Diff}(M)$, the translation part $\theta_{-1}$
corresponding to (the tangent space of) $M$ in the formal quotient (see
introduction) : 
\begin{equation}
 M \simeq \rec{Diff} (M) / \rec{Diff}_x (M) 
 \label{formalquotient}
\end{equation} 
and the $\theta_0 \oplus \theta_1 \oplus \cdots $ part corresponding to the
Maurer-Cartan form on the 'structure group' $\rec{Diff}_x(M)$ of the formally
defined principal bundle 
\begin{equation*}
 \xymatrix{
 \rec{Diff}_x (M) \ar[r] & 
 \rec{Diff}(M) \ar[r] & 
 \rec{Diff} (M) / \rec{Diff}_x (M) \simeq M 
 } 
\end{equation*}

\subsubsection{Structure equations, Bianchi identities} 

On $M_{k+1}$, the $\got{gl}_{-1} \oplus \cdots \oplus \got{gl}_{k-1}$-valued 
2-form 
$$
    \Theta^{k-1} = d \theta^k + \frac{1}{2} [ \theta^k , \theta^k ] \mod
    \got{h}_{k-1}
$$
is tensorial and invariant under $G_{k+1}$, so it descends to a 2-form on $M_k$.
It satisfies the structure equations, analogous to the Maurer-Cartan equations
on a group manifold (recall the formal identification between the frame form and
the Maurer-Cartan form of $\rec{Diff}(M)$) :     
\begin{equation}
     \Theta^{k-1} = d \theta^k + \frac{1}{2} [\theta^k , \theta^k] \mod
     \got{h}_{k-1} = 0
     \label{structure}  
\end{equation}
This is proved in local coordinate form in \cite{koba}, for $k=0,1$. One can
also prove this directly in the same way one proves the Maurer-Cartan equations
for a group. \\
By exterior differentiation of the term $d \theta^k + \frac{1}{2} [\theta^k ,
\theta^k ] $, and use of the structure equations (\ref{structure}), one deduces
the Bianchi type identities : 
\begin{equation}
     \left[ \theta^k , d  \theta^k + \frac{1}{2} [\theta^k, \theta^k] \right] = 0 
     \mod \got{h}_{k-1}  
     \label{bianchi}
\end{equation}
Note that, in contrast with gauge theory, the Bianchi identities are not the
sole consequence of the structure equations.  
          
\subsection{Linear frames : reconstruction of the jet frames}
\label{sectionlinearframes}

We shall denote for later convenience $M = M_{-1}$. 

\subsubsection{1-frames}

A 1-frame above $x \in M$ is an isomorphism 
$$ e_0 : \got{g}_{-1} \to T_x M $$ 
For $e_0$ and $e'_0$ above the same $x$, ${e_0}^{-1} \circ e'_0$ is an
isomorphism of $\got{gl}_{-1}$ so can be written ${e_0}^{-1} \circ e'_0 =
\bad{Ad} (g_0)$ for $g_0 \in GL_0$. So, the space $M_0$ of 1-frames is a
$GL_0$-principal bundle above $M$ with projection $\pi_{0,-1} : 
e_0 \mapsto x$ and right action $e_0 \mapsto e_0. g_0 = R_{g_0}(e_0) = e_0 \circ
\bad{Ad} (g_0)$, which is isomorphic to $M^0$. 
The frame form $\theta = \theta_{-1}$ on $M_0$ is then defined as 
$$ \theta = e_0^{-1} \circ \pi_{0,-1}{_*} $$
It satisfies the same properties as the frame form on $M^0$. 
So, we have a principal bundle structure 
\begin{equation*}
    \xymatrix{
    GL_0 \ar[r] & M_0 \ar[r] & M 
    }
\end{equation*}
such that, at the tangent space level, the following commutative and exact
diagram occurs :
\begin{equation*}
    \xymatrix{
    \got{gl}_0 \ar[r] \ar[d] & 
    \got{gl}_0 \oplus \got{gl}_{-1} \ar[r] &
    \got{gl}_{-1} \ar[d]^-{e_0} \\
    T_0 M_0 \ar[r] & 
    T_{e_0} M_0 \ar[r]^{\pi_{0,-1}{_*}} \ar[ru]^-{\theta_{-1}} & 
    T_x M 
    }
\end{equation*}      
Note the well known fact \cite{sharpe,cap} that this is this last point which makes the difference
between gravity and ordinary gauge theory.  
\subsubsection{$k$-frames, $k > 1$}

{\bf{Induction hypothesis}}
\\ \\
Assume now we have constructed spaces $M_{l}$ of $e_l$'s, for $0 \leq l \leq k$,
which are isomorphic to the $M^l$, and so have the same structure and same
properties as displayed previously. We denote $\pi_{k,l-1} : M_k \to M_{l-1}$
the projections, and $T_l M_k = \ker \pi_{k,l-1}{_*}$. 
We shall construct the space $M_{k+1}$
isomorphic to $M^{k+1}$ by a prolongation procedure similar to those of
\cite{koba}, \cite{cap}. 
\\ \\
{\bf First prolongation of $M_k$}
\\ \\
We define a $(k+2)$-frame above $e_k \in M_k$ as an isomorphism 
$$ 
e_{k+1} : \got{gl}_{-1} \oplus \cdots \oplus \got{gl}_k \to T_{e_k} M_{k}
$$
such that the following diagram commute :
\begin{equation*}
    \xymatrix{
    \got{gl}_0 \oplus \cdots \got{gl}_k \ar[r] \ar[d]^{\hat{}} & 
    \got{gl}_{-1} \oplus \cdots \oplus \got{gl}_k \ar[r] \ar[d]^-{e_{k+1}} &
    \got{gl}_{-1} \oplus \cdots \oplus \got{gl}_{k-1} \ar[d]^-{e_k} \\
    T_0 M_k \ar[r] & T_{e_k} M_k \ar[r] & T_{e_{k-1}} M_{k-1}
    }
\end{equation*}
Let $M_{k,1}$ be the space of the $e_{k+1}$'s.
\\ \\
{\bf Principal bundle structure}
\\ \\
$ \bullet $
For $e_{k+1}$ and $e'_{k+1}$ above the same $e_k$, the definition then implies
that the isomorphism ${e_{k+1}}^{-1} \circ e'_{k+1}$ of $\got{gl}_{-1} \oplus
\cdots \oplus \got{gl}_k$ is of degree $k$ i.e. 
$$ {e_{k+1}}^{-1} \circ e'_{k+1} = \bad{Ad} (g_{k,1}) , \; g_{k,1} \in GL_{k,1} $$
Alternatively, this means we have constructed, above $e_k \in M_k$, the
commutative square : 
\begin{equation*}
    \xymatrix{
    \got{gl}_{-1} \oplus \cdots \got{gl}_k \ar[r]^-{ \bad{Ad}(g_{k,1}) } 
     \ar[d]^-{e'_{k+1}} & 
    \got{gl}_{-1} \oplus \cdots \got{gl}_k \ar[d]^-{e_{k+1}} \\
    T_{e_k} M_k \ar[r] & T_{e_k} M_k 
    }
\end{equation*}          
All this proves that the projection $\pi_{k+1,k} : e_{k+1} \mapsto e_{k}$, and
right action $e_{k+1} \mapsto e_{k+1} \circ \bad{Ad} (g_{k,1})$ identifie the
principal bundle :
\begin{equation}
     \xymatrix{
     GL_{k,1} \ar[r] & M_{k,1} \ar[r] & M_{k} 
     }
     \label{fibration1}
\end{equation}
$ \bullet $
Next, consider $e_{k+1}$, $e'_{k+1}$ above the same $x \in M$ for the projection
$\pi_{k+1,-1} = \pi_{k,-1} \circ \pi_{k+1,k}$. Then $e_{k+1}$, $e'_{k+1}$ are
above $e_k$, $e'_k$ with $e'_k = e_k . g^k$, $g^k \in GL^k$.      
For any $g^{k+1}$ above $g^k$, with respect to the projection $GL^{k+1} \to
GL^{k}$, we define $e''_{k+1} = R_{g^k}{_*} e_{k+1} \circ \bad{Ad} (g^{k+1})$
(see equation (\ref{transfjetk})).
Then $e''_{k+1}$ is a $(k+2)$-linear frame above $e'_k$.
\\
So, by the preceding point, we have $g_{k,1} \in GL_{k,1}$ such that $e''_{k+1}
\circ \bad{Ad} (g_{k,1}) = e'_{k+1}$, and we obtain 
\begin{equation}
     e'_{k+1} = R_{g^k}{_*} e_{k+1} \circ \bad{Ad} (g^{k,1}) 
     \label{rightaction}
\end{equation}
with $ g^{k,1} = g^{k+1} . g_{k,1} $               
In one word, we have just constructed the commutative squares : 
\begin{equation*}
    \xymatrix{
    \got{gl}_{-1} \oplus \cdots \oplus \got{gl}_k \ar[r]^-{\bad{Ad}g_{k,1}}
    \ar[d]^-{e'_{k+1}} & 
    \got{gl}_{-1} \oplus \cdots \oplus \got{gl}_k \ar[r]^-{\bad{Ad}g^{k+1}}
    \ar[d]^-{e''_{k+1}} & 
    \got{gl}_{-1} \oplus \cdots \oplus \got{gl}_k \ar[d]^-{e_{k+1}} \\
    T_{e'_k} M_k \ar[r]^-{\rec{id}} & 
    T_{e'_k} M_k \ar[r]^-{R_{g^k}{_*}^{-1}} & 
    T_{e_k} M_k 
    }
\end{equation*}    
Thus, the projection $\pi_{k+1,-1}$ and the right action identifie $M_{k,1}$ as a
$GL^{k,1}$-principal bundle above $M$ : 
\begin{equation}
       \xymatrix{
       GL^k \ltimes GL_{k,1} = GL^{k,1} \ar[r] & 
       M_{k,1} \ar[r] & 
       M 
       }
       \label{fibration2}
\end{equation}
$ \bullet $
The principal fibrations (\ref{fibration1}) and (\ref{fibration2}), are
summarised in  
\begin{equation*}
   \xymatrix{
   GL_{k,1} \ar[r] \ar@{=}[d] & 
   GL^{k,1} = GL^k \ltimes GL_{k,1} \ar[r] \ar[d] & 
   GL^k \ar[d] \\
   GL_{k,1} \ar[r] & 
   M_{k,1} \ar[r]^-{\pi_{k+1,k}} \ar[d]^{\pi_{k+1,-1}} & 
   M_k \ar[d]^-{\pi_{k,-1}} \\
   & 
   M \ar@{=}[r] & M 
   }
\end{equation*}   
{\bf Frame form}
\\ \\
On $M_{k,1}$, we define the frame form $\theta^{k}$ as : 
\begin{equation*}
     \theta^{k} = {e_{k+1}}^{-1} \pi_{k+1,k}{_*}
\end{equation*}
Then, the definition of right action (\ref{rightaction}) and definition of
$(k+2)$-frames are dually encoded in the properties : 
\begin{eqnarray*}
    &(i)& R_{g}{^*} \theta^k = \bad{Ad} (g^{-1}) \theta^k , \; g \in GL^{k,1} \\
    &(ii) & \theta^k (\hat X_0 \oplus \cdots \oplus \hat X_{k+1}) = X_0
    \oplus \cdots X_{k} , \,\,\, \ker \theta^k = \ker \pi_{k+1,k}{_*} = T_{k+1}
    M^{k+1} \\
    &(iii)& \pi_{k+1,k}^{*} \theta^{k-1} = \theta^k \mod \got{gl}_k 
\end{eqnarray*}
From this, we define the curvature form as :
\begin{equation}
    \Theta^{k-1} = d \theta^k + \frac{1}{2} [\theta^k, \theta^k] \mod
    \got{h}_{k-1} 
\end{equation}
Horizontality of the frame form $(ii)$ then proves $\Theta^{k-1}$ is basic,
$i_{\hat X} \Theta^{k-1} = 0$, $X = X_0 \oplus \cdots \oplus X_{k,1}$.
Equivariance $(i)$ proves that $\Theta^{k-1}$ is equivariant under $GL^k$ : 
\begin{equation}        
      R_{g^k}^* \Theta^{k-1} = \bad{Ad} (({g^k})^{-1}) \Theta^{k-1} 
\end{equation}
and transforms affinely under $GL_{k,1}$ : 
\begin{equation*}
      R_{g_{k,1}}^* \Theta^{k-1} = \Theta^{k-1} - \partial \alpha_k \circ
      \theta_{-1} 
\end{equation*}
Finally, the recursive property $(iii)$ and induction hypothesis prove the
recursive identity :
\begin{equation}
     \pi_{k+1,k}^* \Theta^{k-2} = \Theta^{k-1} \mod \got{gl}_{k-1} = 0
\end{equation}
All the properties of $\Theta^{k-1}$ are then equivalently encoded in the
torsion map 
$$ t : M_{k,1} \to \got{gl}_{k-1,2} 
       = \got{gl}_{k-1} \otimes \Lambda^2 \got{gl}_{-1}^* $$
which maps $e_{k+1}$ to $t_{e_{k+1}}$ with : 
\begin{equation*}
     t_{e_{k+1}} (X_{-1}, Y_{-1}) = \Theta^{k-1} (\overline{e_{k+1}(X_{-1})},
                                    \overline{e_{k+1}(Y_{-1})}) 
				  = d (\theta^{k-1})_{k-1} (e_{k+1}(X_{-1}).
				  e_{k+1} (Y_{-1})) 
\end{equation*}
where $\overline{e_{k+1}(X_{-1})}$ is a lift of $e_{k+1}(X_{-1}) \in T M_k$ to
$T M_{k,1}$, and where $(\theta^{k-1})_{k-1}$ is the component of degree $(k-1)$
of the frame form $\theta^{k-1}$ on $M_k$. We then have the covariance
properties : 
\begin{equation}
     t_{e_{k+1}.g^k} = \bad{Ad} ({g^{k}}^{-1}) \circ t_{e_{k+1}} \circ \bad{Ad}
     (g_0) \,\,\, , \,\,\, 
     t_{e_{k+1}.g_{k,1}} = t_{e_{k+1}} - \partial \alpha_k		            
     \label{equivartorsion}
\end{equation}
We summarise this by saying the following diagram is commutative and
covariant under the $GL^k$ action : 
\begin{equation*}
     \xymatrix{
     GL_{k,1} \simeq \got{gl}_{k,1} \ar[r] \ar[d]^-{\partial} & M_{k,1} \ar[d]^-t \\
     \got{gl}_{k-1,2} \ar@{=}[r] & \got{gl}_{k-1,2} 
     }
\end{equation*}     
\\ \\
{\bf Reduction to $M_k$}
\\ \\                   
Now, by evaluating the Bianchi identities of $M_{k}$ (satisfied by the induction
hypothesis) 
$$
     \left[ 
     \theta^{k-1} , d \theta^{k-1} + \frac{1}{2} [\theta^{k-1} , \theta^{k-1}] 
     \right]
     = 0 \mod \got{h}_{k-2} 
$$
on vectors $e_{k+1} (X_{-1}), e_{k+1} (Y_{-1}),e_{k+1} (Z_{-1})$, we obtain 
$$
     \partial t_{e_{k+1}}  = 0 
$$
so that the torsion at each $e_{k+1}$ is a $\partial$-cocyle. This last property
and the exactness of the $\partial$-sequence 
\begin{equation}
     \xymatrix{
     \got{gl}_{k+1} \ar[r] & 
     \got{gl}_{k,1} \ar[r]^-{\partial} & 
     \got{gl}_{k-1,2} \ar[r]^-{\partial} & 
     \got{gl}_{k-2,3}  
     }
     \label{sequencepartial}
\end{equation}
at $\got{gl}_{k-1,2}$ then proves that we have $t_{e_1} = \partial \alpha_k$,
for a $\alpha_k \in \got{gl}_{k,1} \simeq GL_{k,1}$. Thanks to equivariance  
(\ref{equivartorsion}), all this proves the existence of
$(k+2)$-frames with null torsion, i.e. the map $t$ has a kernel. We then simply
define 
$$ 
    M_{k+1} = t^{-1} (0)
$$
that is $M_k$ are the $e_{k+1}$ such that $t_{e_{k+1}} = 0$. Then both the 
equivariance 
(\ref{equivartorsion}) and the exactness of (\ref{sequencepartial}) at
$\got{gl}_{k,1}$ then prove that $M_{k+1} \to M_k$ is a subbundle of $M_{k,1}
\to M_k$ with structure group $GL_{k+1} \simeq \got{gl}_{k+1}$. All these facts
are summarised in the exact commutative diagram, which completes the diagram
following equation (\ref{equivartorsion}) :       
\begin{equation*}
  \xymatrix{
  \got{gl}_k \simeq GL_k \ar[r] \ar[d] & 
  M_{k+1} \ar[r] \ar[d] & 
  M_k \ar@{=}[d] \\
  \got{gl}_{k,1} \simeq GL_{k,1} \ar[r] \ar[d]^-{\partial} &
  M_{k,1} \ar[r] \ar[d]^-t & 
  M_k \\
  \got{gl}_{k-1,2}  \ar@{=}[r] \ar[d]^-{\partial}
  & 
  \got{gl}_{k-1,2} \ar[d]^-{\partial} \\
  \got{gl}_{k-2,3} \ar@{=}[r] & 
  \got{gl}_{k-2,3}  
  }
\end{equation*}  
The first column describes an exact Spencer $\partial$-sequence, the second the
construction of $M_{k+1}$, and the first two lines the principal fibrations so obtained.  
\\ \\
{\bf Structure of $M_{k+1}$}
\\ \\
We have thus obtained an iterative fibration 
\begin{equation*}
   \xymatrix{
   GL_{k+1} \ar[r] \ar@{=}[d] & 
   GL^{k+1} \ar[r] \ar[d] & 
   GL^{k} \ar[d] \\
   GL_{k+1} \ar[r] & 
   M_{k+1} \ar[r] \ar[d] & 
   M_k \ar[d] \\
   & 
   M \ar@{=}[r] & M 
   }
\end{equation*}   
$M_{k+1}$ is equiped with the frame form $\theta^{k}$ inherited from $M_{k,1}$,
and now we have, as the torsion of $e_{k+1}$ vanish :
\begin{equation*}
    \Theta^{k-1}  = 0
\end{equation*}
i.e. the structural equations. So $M_{k+1}$ has the same structure as $M_{k}$ at
the next order.
Using the induction hypothesis $M_k \simeq M^k$, the map 
$$ 
    e^{k+1} \mapsto e_{k+1} 
$$
defined in section \ref{sectioninterpretation}, is then, by construction, an
isomorphism of principal bundles, so $M_{k+1} \simeq M^{k+1}$.    

\section{Field theory of frames}

\subsection{Preliminaries : Local fields} 

\subsubsection{Local Spencer cochains}
\label{sectionlocalspencercochains}

$\bullet$
To order $k+2$, one obtains a local version of $\got{gl}_{-1} \oplus \cdots
\oplus \got{gl}_k$ by defining the associated bundle  
\begin{equation}
     S_k = M_{k+1} \times_{\bad{Ad}} \got{gl}_{-1} \oplus \cdots \oplus \got{gl}_k
\end{equation}
which can be seen as a higher order tangent bundle above $M$. 
Local Spencer cochains are $S_k$-valued forms on $M$, i.e. elements of $\Omega^*
(M,S_k)$. These are the basic fields of the theory.  
\\ \\ 
$\bullet$ 
Owing to the structure of $M_{k+1}$, we can give alternative and useful descriptions of
this. First, recall we have 
$$
    \Omega^l (M,S_k) \simeq \Omega^l_{GL^{k+1}} (M_{k+1},\got{gl}_{-1} \oplus \cdots
    \oplus \got{gl}_k) 
$$
Second, this isomorphy allows to associate to each $\alpha \in \Omega^l(M,S_k)$
the function $\tilde{\alpha}$ on $M_{k+1}$ defined at each $e_{k+1}$ by 
$$
     \tilde{\alpha}_{|e_{k+1}} (X_1 , \cdots X_l) 
     = \alpha_{|e_{k+1}} (e_{k+2}(X_1) , \cdots , e_{k+2} (X_l)) 
     \Leftrightarrow 
     \alpha = \tilde{\alpha} \circ \theta_{-1} 
$$
for any $e_{k+2}$ above $e_{k+1}$, $X_i \in \got{gl}_{-1}$. 
We shall extend each $\tilde{\alpha}_{|e_{k+1}}$ to a null form on $\got{gl}_0
\oplus \cdots \oplus \got{gl}_k$, so that we will also write $\alpha =
\tilde{\alpha} \circ \theta$.  
As $\tilde{\alpha}$ is then equivariant, this naturally 
defines an isomorphy between $\Omega^l (M,S_k)$ and the space of section of the
bundle
$$
     M_{k+1} \times_{\bad{Ad}} (\got{gl}_{-1} \oplus \cdots \oplus \got{gl}_k) 
     \otimes \Lambda^l \got{gl}_{-1}^* 
$$
This last fact implies that we can define, point by point, an algebraic $\partial$-operator on
$\Omega^l(M,S_k)$. Third, to $\alpha$ one can also associate the vector valued
form $\bar{\alpha}$ defined as  
$$ 
    \bar{\alpha}_{|e_{k+1}} (u_1, \cdots , u_l ) = e_{k+1} (\alpha_{|e_{k+1}} 
    (u_1,\cdots, u_l)) 
$$
for $u_i \in TM_{k+1}$. 
This means $\Omega^l (M,S_k)$ is also isomorphic with the tensorial forms on
$M_{k+1}$ with values in tangent vector on $M_k$. Then, as on any space of Lie
algebra valued forms, we can define the standard structure of differential graded Lie
algebra, thus obtaining the algebraic as well as differential brackets of
\cite{pom}. 

\subsubsection{Linear Spencer sequences}

For $\alpha \in \Omega^l (M,S_k)$, viewed as a tensorial form on $M_{k+1}$, we
define :
\begin{equation*}
     d_{\theta} \alpha = d \alpha + [ \theta , \alpha ] \mod \got{h}_{k-1} 
\end{equation*}
Then $d_{\theta} \alpha$ is still tensorial, and this defines a map 
\begin{equation*} 
     d_{\theta} : \Omega^l (M,S_k) \to \Omega^{l+1} (M,S_{k-1})        
\end{equation*}
The structure equation $\Theta = 0$ on $M_{k+1}$, then proves that $d_{\theta}$
is nilpotent 
\begin{equation*}
     d_{\theta}^2 \alpha = [\Theta , \alpha ] \mod \got{h}_{k-2} = 0 
\end{equation*}
thus giving the linear sequence
\begin{equation}    
    \xymatrix{ 
    \Omega^0 (M,S_k) \ar[r]^-{d_{\theta}} &
    \Omega^1 (M,S_{k-1}) \ar[r]^-{d_{\theta}} &
    \cdots \ar[r] & 
    \Omega^n (M,S_{k-n}) \ar[r]^-{d_{\theta}} &
    0 
    }
    \label{linearspencersequence1}
\end{equation}
The proof of this is a straightforward application of the definitions. 
In the following, we shall complete this sequence to the linear Spencer
sequence.     
\subsection{Symmetries}

\subsubsection{Diffeomorphisms} 

We denote by $\rec{Aut}(M)$ the group of (oriention preserving) diffeomorphisms 
of $M$. Let $f = f_{-1} \in \rec{Aut}(M)$. 
\\
From the jet viewpoint, $f$ acts on $M^k$ by 
$$
    e^k = j^{k+1}_0 (\phi) \mapsto f_k(e_k) = j^{k+1}_0 (f \circ \phi)
$$    
Let us analyse this from the linear frame viewpoint. 
The action on $M_0$ is given by 
$$ e_0 \to f_0(e_0) = f_{-1}{_*} e_0 $$
Then $f_0$ satisfies $R_{g_0} \circ f_0 = f_0 \circ R_{g_0}$, $g_0 \in GL_0$,
and $\pi_{0,-1} \circ f_0 = f_{-1} \circ \pi_{0,-1}$, so is a principal bundle
automorphism. Moreover, we have 
\begin{eqnarray*}
     {f_0^* \theta^{-1}}_{|e_0}
     &=& 
     {\theta^{-1}}_{|f_0(e_0)} \circ f_0{_*} 
     = 
     {f_0(e_0)}^{-1} \pi_{0,-1}{_*}f_0{_*} \\
     &=&
     {f_0(e_0)}^{-1} f_{-1}{_*} \pi_{0,-1}{_*} = 
     {e_0}^{-1} \pi_{0,-1}{_*} \\
     &=&
     {\theta^{-1}}_{|e_0}
\end{eqnarray*}     
This shows that the action on $M_1$ defined by 
$$ e_1 \to f_1(e_1) = f_0{_*} e_1 $$
is well defined (i.e. $e_1$ is a 2-frame of null torsion).
Recursively, we define $f_{k+1}$ from $f_k$ by :
$$ f_{k+1} (e_{k+1}) = f_k{_*} e_{k+1} $$ 
Exactly the same calculation as before proves this is well defined.   
Then the prolongated diffeomorphisms satisfies :
$$ R_{g^k} \circ f_k = f_k \circ R_{g^k} , \; 
   \pi_{k,k-1} \circ f_k = f_{k-1} \circ \pi_{k,k-1}
$$
and keep invariant the frame form (same calculation as for $\theta^{-1}$)
\begin{equation}
    {f_{k}}^* \theta^{k-1} = \theta^{k-1} 
\end{equation}    
We shall denote $j_k(f)=f_k$ the prolongated diffeomorphism. 

\subsubsection{Extended diffeomorphisms} 

$\bullet$ 
Now, denote by $\rec{Aut}(M_{k})$ the automorphism group of $M_k \to M$ as a
principal fiber bundle, that is : 
$$ f_k \in \rec{Aut} (M_k) \; : \; 
   f_k \circ R_{g^k} = R_{g^k} \circ f_k 
$$
Then $\rec{Aut}(M_k)$ is a a subgroup of the group of diffeomorphisms of $M_{k}$
which preserves the fibers of $M_k \to M$.       
The gauge group ${\cal{GL}}^k$ of $M_k \to M$ are the vertical automorphisms in
$\rec{Aut}(M_k)$ i.e. 
$$
    f_k \in {\cal{GL}}^k : f_k \circ R_{g^k} = R_{g^k} \circ f_k 
    \,\,\, , \,\,\,  
                           f_k (\pi_{k,-1}^{-1} (x)) = \pi_{k,-1}^{-1} (x) 
$$
for all $x \in M$. 
Similarly, we define the gauge group ${\cal{GL}}_{k}$ of $M_{k} \to M_{k-1}$,
and we observe that the gauge group of $M_{k} \to M_{k'}$, $k' \leq k$, is $
{\cal{GL}}_{k'+1} \ltimes \cdots \ltimes {\cal{GL}}_k$, so that in particular 
$$
    {\cal{GL}}^{k+1} \simeq {\cal{GL}}_0 \ltimes \cdots \ltimes {\cal{GL}}_{k} 
$$    			   
As usual, gauge transformations, in ${\cal{GL}}^k$ say, are isomorphic with
section of the adjoint bundle $M_{k} \times_{\rec{Ad}} GL^k$, thanks to the
isomorphy $g^{k} \mapsto \tilde{g}^{k}$ defined by 
$$
     g^{k} (e_{k}) = e_k . \tilde{g}^k (e_{k}) = R_{\tilde{g}^k(e_k)} (e_k) 
$$       
\\
$\bullet$ 
We define projections $f_{k+1} \mapsto f_{k}$ by 
$$
    f_{k} (e_{k}) = \pi_{k+1,k} (f_{k+1}(e_{k+1})) 
$$
for any $e_{k+1}$ above $e_{k}$. This is well defined thanks to the equivariance
of $f_{k+1}$, and we have $\pi_{k+1,k} \circ f_{k+1} = f_k \circ \pi_{k+1,k}$.
In other words, we have commutation in  
\begin{equation*}
     \xymatrix{
     GL_{k+1} \ar[r] \ar[d] & 
     M_{k+1} \ar[r]^-{\pi_{k+1,k}} \ar[d]^-{f_{k+1}} &
     M_{k} \ar[d]^-{f_k} \\
     GL_{k+1} \ar[r] &
     M_{k+1} \ar[r]^-{\pi_{k+1,k}} & 
     M_k 
     }
\end{equation*}       
and we obtain the tower of commutative squares
\begin{equation*}
    \xymatrix{ 
    M_{k+1} \ar[r] \ar[d]^-{f_{k+1}} & 
    M_k \ar[r] \ar[d]^-{f_k} & 
    \cdots \ar[r] & 
    M_0 \ar[r] \ar[d]^-{f_0} & 
    M \ar[d]^-{f_{-1}} \\
    M_{k+1} \ar[r] & 
    M_k \ar[r] & 
    \cdots \ar[r] & 
    M_0 \ar[r] & 
    M
    }
\end{equation*}    
Note that these projections are group morphisms from 
$\rec{Aut}(M_{k+1})$
to $\rec{Aut}(M_{k})$. For $f_{k+1}$, $f'_{k+1}$ projecting on the same $f_{k}$, the
automorphism $f''_{k+1} = {f_{k+1}}^{-1} \circ f'_{k+1}$ then preserves the
fibers of $M_{k+1} \to M_{k}$, and is thus a gauge transformation : 
$$ f'_{k+1} = f_{k+1} \circ f''_{k+1} \,\,\, , \,\,\, 
   f''_{k+1} \in {\cal{GL}}_{k+1} $$
So we obtain a principal bundle 
\begin{equation*}
    \xymatrix{
    {\cal{GL}}_{k+1} \ar[r] & \rec{Aut} (M_{k+1}) \ar[r] & 
    \rec{Aut} (M_{k}) 
    }
\end{equation*}
with gauge transformations projecting on the identity of $\rec{Aut}(M_k)$. 
More generally we obtain in this way principal bundles : 
\begin{equation*}
     \xymatrix{
    {\cal{GL}}_{k'+1} \ltimes \cdots \ltimes {\cal{GL}}_k \ar[r] & 
    \rec{Aut} (M_{k}) \ar[r] & 
    \rec{Aut} (M_{k'}) 
    }
\end{equation*}
and in particular 
\begin{equation}
    \xymatrix{
    {\cal{GL}}^{k+1} \ar[r] & \rec{Aut} (M_{k+1}) \ar[r] & 
    \rec{Aut} (M)
    }
    \label{principalprincipal}
\end{equation}
This last bundle admits the global section given by 
$f_{-1} \to f_{k+1} = j_{k+1} (f_{-1})$. The section $j_{k+1}$ enables us to
construct, for $f_{k+1} \in \rec{Aut}(M_{k+1})$ projecting on $f_{-1} \in
\rec{Aut}(M)$, the gauge transformation $g^{k+1} \in
{\cal{GL}}^{k+1}$ defined by : 
\begin{equation}
     f_{k+1} = j_{k+1}(f_{-1}) \circ g^{k+1} 
     \label{diffversusgauge}
\end{equation}
The equation (\ref{diffversusgauge}) gives a global trivialization of
(\ref{principalprincipal}), that is of the semi-direct product 
$     
     \rec{Aut} (M_{k+1})  \simeq  \rec{Aut}(M) \ltimes {\cal{GL}}^{k+1} 
$.                   
\subsubsection{Synthesis}

For $f_{k+1} \in \rec{Aut} (M_{k+1})$, we define the first Spencer operator 
$$ D_{\theta} f_{k+1}  = {f_{k+1}}^* \theta^k  - \theta^k $$
Then $D_{\theta} f_{k+1} $ is a tensorial $(\got{gl}_{-1} \oplus \cdots \oplus
\got{gl}_k)$-valued 1-form, that is $D_{\theta} f_{k+1} \in \Omega^1(M,S_k)$. Indeed, for $X = X_{0} \oplus \cdots \oplus
X_{k+1}$, the equivariance of $f_{k+1}$ implies : 
\begin{equation*}
     i_{\hat X} D_{\theta} f_{k+1} = \theta^{k} (f_{k+1}{_*} \hat X) - \theta^{k}
     (\hat X) = \theta^k (\hat X) - \theta^k (\hat X) = 0 
\end{equation*}
and, for $g \in GL^{k+1}$,  
\begin{eqnarray*}
    R_{g}^* D_{\theta} f_{k+1} &=& R_{g}^* {f_{k+1}}^* \theta^{k} 
        - R_{g}^* \theta^k = 
	{f_{k+1}}^* R_{g}^* \theta^k - R_{g}^* \theta^{k} 
	\\
	&=& \bad{Ad} (g^{-1}) {f_{k+1}}^{*} \theta^{k} - \bad{Ad} (g^{-1})
	\theta^{k}  =  
	\bad{Ad} (g^{-1}) D_{\theta} f_{k+1}  
\end{eqnarray*}	      
Moreover, \\\\
{\it{
$D_{\theta}$ defines a cocycle on the group $\rec{Aut} (M_{k+1})$ with values in
$\Omega^1 (M,S_k)$, with kernel the group of diffeomorphisms of $M$, that is
$D_{\theta} f_{k+1}  = 0$ iff $f_{k+1} = j_{k+1}(f_{-1})$.  
}} 
\\ \\ 
Indeed, the cocycle relation follows from : 
\begin{eqnarray*}
     D_{\theta} (f_{k+1} \circ g_{k+1}) 
     &=& 
     (f_{k+1} \circ g_{k+1})^* \theta^k - \theta^k
     = 
     {g_{k+1}}^* {f_{k+1}}^* \theta^k - \theta^k \\
     &=&
     {g_{k+1}}^* \left( {f_{k+1}}^* \theta^k - \theta^k \right)
     + {g_{k+1}}^* \theta^k - \theta^k \\
     &=& 
     {g_{k+1}}^* D_{\theta} f_{k+1}  + D_{\theta} g_{k+1}  
\end{eqnarray*}
Next, we have already shown that $f_{k+1} = j_{k+1}(f)$, $f \in \rec{Aut}(M)$, keeps the frame form
invariant, i.e. $D_{\theta} f_{k+1} = 0$. Conversely, suppose $D_{\theta} f_{k+1} = 0$ i.e.
${f_{k+1}}^* \theta^k = \theta^k$. As 
\begin{eqnarray*}
     {f_{k+1}}^* {\theta^k}_{|e_{k+1}} 
     & = & 
     {\theta^k}_{|f_{k+1}(e_{k+1})} \circ f_{k+1}{_*} \\
     & = & 
     {f_{k+1}(e_{k+1})}^{-1} \pi_{k+1,k}{_*} f_{k+1}{_*} \\
     & = & 
     {f_{k+1}(e_{k+1})}^{-1} f_k {_*} \pi_{k+1,k}{_*}
\end{eqnarray*}
the equation ${f_{k+1}}^* \theta^k = \theta^k$ implies 
$
     {f_{k+1}(e_{k+1})}^{-1} f_k{_* } \pi_{k+1,k}{_*} = {e_{k+1}}^{-1}
     \pi_{k+1,k}{_*} 
$ 
and so $D_{\theta} f_{k+1} = 0 $ is equivalent to : 
\begin{equation}
    f_{k+1} (e_{k+1}) = f_k{_*} e_{k+1}
    \label{equalityframe} 
\end{equation}                
Then, from equation (\ref{equalityframe}) the result is easily proved by
induction on $k$.
\begin{flushright} $\blacksquare$ \end{flushright}   
All this is summarised in the exact sequence
\begin{equation}
     \xymatrix{ 
     \rec{id} \ar[r] & 
     \rec{Aut}(M) \ar[r]^-{j_{k+1}} & 
     \rec{Aut}(M_{k+1}) \ar[r]^-{D_{\theta}} & 
     \Omega^1 (M,S_k) \ar[r] & 
     0 
     }
     \label{spencer11}
\end{equation}          		 
$j_{k+1}$ being a group morphism and $D_{\theta}$ a group cocycle. 

\subsubsection{Action of $\rec{Aut}(M_{k+1})$ on local fields}
\label{actionofextdiffeoonlocalfields}

It is useful for next purpose to compute the action of an extended
diffeomorphism on a local field. 
\\ \\
$\bullet$ As a preliminary, take $f_{k+1} \in \rec{Aut}(M_{k+1})$, then as
$D_{\theta} f_{k+1} \in
\Omega^1(M,S_k)$, we can view it as a function $\tilde{D}_{\theta} f_{k+1}$ on $M_{k+1}$
with values in $\got{gl}_{-1} \oplus \cdots \oplus \cdots \got{gl}_k$ (section 
\ref{sectionlocalspencercochains}). For $X = X_{-1} \oplus \cdots \oplus
X_k$, and any $e_{k+2}$ above $e_{k+1}$, one finds : 
\begin{eqnarray*}
     {\tilde{D}_{\theta} f_{k+1} }_{|e_{k+1}} (X) 
     & = & 
     {D_{\theta} f_{k+1}}_{|e_{k+1}} (e_{k+2} (X)) \\ 
     & = & 
     {{f_{k+1}}^* \theta}_{|e_{k+1}} (e_{k+2} (X)) - X \\
     & = & 
     {f_{k+1}(e_{k+1})}^{-1} \pi_{k+1,k}{_*} f_{k+1}{_*} e_{k+2} (X) - X 
\end{eqnarray*}
All this proves that the map 
$$ X \mapsto X + {f_{k+1}(e_{k+1})}^{-1} \pi_{k+1,k}{_*} f_{k+1}{_*} e_{k+2} (X)
$$
that we shall denote $ 1 + \tilde{D}_{\theta} f_{k+1}$, is an automorphism of
$\got{gl}_{-1} \oplus \cdots \oplus \got{gl}_k$, inducing the identity on the
$\got{gl}_0 \oplus \cdots \oplus \got{gl}_k$ part, and that this is indeed the
equivariant version of $ \theta + D_{\theta} f_{k+1} = {f_{k+1}}^* \theta $. 
\\ \\ 
$\bullet$ Take a local field $\alpha \in \Omega^*(M,S_k)$. Viewing $\alpha$ as 
a tensorial form on $M_{k+1}$, the action of
$f_{k+1}$ is simply 
$$  \alpha \to \alpha' = {f_{k+1}}^* \alpha $$
Equivariance of $f_{k+1}$ shows this is consistent. 
\\ \\
$\bullet$ View now $\alpha$ as a equivariant function $\tilde{\alpha}$ on
$M_{k+1}$ (section \ref{sectionlocalspencercochains}). $\tilde{\alpha}$ 
transforms to
$\tilde{\alpha}'$. At $e_{k+1}$, we have 
\begin{eqnarray*}
     {\tilde{\alpha}' \circ \theta}_{|e_{k+1}} 
     & = & 
     \alpha'_{|e_{k+1}} = {{f_{k+1}}^* \alpha}_{|e_{k+1}} \\
     & = & 
     \alpha_{|f_{k+1}(e_{k+1})} \circ f_{k+1}{_*}{_{|e_{k+1}}} \\
     & = & 
     \tilde{\alpha}_{|f_{k+1}(e_{k+1})} \circ \theta_{|f_{k+1}(e_{k+1})} 
     \circ f_{k+1}{_*}{_{|e_{k+1}}} 
\end{eqnarray*}
and, by evaluating on $e_{k+2}(X)$, for any $e_{k+2}$ above $e_{k+1}$ : 
\begin{eqnarray*}
     \tilde{\alpha}'_{|e_{k+1}} (X) 
     & = & 
     \tilde{\alpha}_{|f_{k+1}(e_{k+1})}
     (\theta_{|f_{k+1}(e_{k+1})}(f_{k+1}{_*}e_{k+2} (X))) \\ 
     & = & 
     \tilde{\alpha}_{|f_{k+1}(e_{k+1})} (1 + \tilde{D}_{\theta} f_{k+1})_{|e_{k+1}} 
\end{eqnarray*}
All this means that, from the equivariant viewpoint, the field $\tilde{\alpha}$
transforms as : 
\begin{equation}
     \tilde{\alpha} \to \tilde{\alpha}' = 
     {f_{k+1}}^* \tilde{\alpha} \circ (1 + \tilde{D}_{\theta} f_{k+1} )
     \label{actionextendeddiffeo}
\end{equation}
Note that if $f_{k+1}$ comes from a diffeomorphism, i.e. $D_{\theta} f_{k+1} = 0$, then
the preceding tranformation law is, as expected, $\tilde{\alpha} \to {f_{k+1}}^*
\tilde{\alpha}$. 
               
\subsection{Deformations}

\subsubsection{Deformation space}
\label{sectiondeformationspace}

$\bullet$ 
Consider the group $B_{k}$ of automorphisms of $\got{gl}_{-1} \oplus \cdots \oplus
\got{gl}_k$ which induce the identity on $\got{gl}_0 \oplus \cdots \oplus
\got{gl}_k$. This group then consists in inversible transformations such that : 
\begin{equation*}
    X_{-1} \oplus X_0 \oplus \cdots \oplus X_k 
    \mapsto 
    X_{-1} + \mu_{-1} (X_{-1}) \oplus X_0 + \mu_0 (X_{-1}) \oplus \cdots 
    \oplus X_k + \mu_k (X_{-1}) 
\end{equation*}
where $\mu_l \in \got{gl}_{l,1}$. We denote this simply $1 + \tilde{\mu}$. 
The inverse transformation is 
\begin{eqnarray*}
     X_{-1} \oplus \cdots X_0 \oplus \cdots \oplus X_{k} 
     & \mapsto & 
     \left( 1 + \tilde{\mu}_{-1} \right)^{-1} X_{-1} \oplus 
     X_0 - \tilde{\mu}_0 \left( 1 + \tilde{\mu}_{-1} \right)^{-1} X_{-1} 
     \oplus \\
     &  & \,\,\, 
     \cdots \oplus 
     X_k - \tilde{\mu}_k \left( 1 + \tilde{\mu}_{-1} \right)^{-1} X_{-1}
\end{eqnarray*}
so that $1+\tilde{\mu} \in B_k$ iff $1+\mu_{-1} \in B_{-1} = GL_0$. As
$B_k$ is a subspace of $\got{gl}_{-1,1} \oplus \cdots \oplus \got{gl}_{k,1}$, 
$GL^{k+1}$ acts on the left with $\bad{Ad}$ on it (preserving the inversibility
property), and we can define the associated fiber bundle 
$$ {\cal{B}}_k = M_{k+1} \times_{\bad{Ad}} B_{k} $$
To each section $\tilde{\mu} \in \Gamma ({\cal{B}}_k)$ seen as a equivariant
$B_k$-valued function on $M_{k+1}$, we can associate the tensorial 
one-form $\mu \in \Omega^1 (M,S_k)$ defined by (see section
\ref{sectionlocalspencercochains}) : 
\begin{equation}
    \mu = \tilde{\mu} \circ \theta_{-1} 
    \label{iso1}
\end{equation} 
so that we have the identity 
$$ (1 + \tilde{\mu}) \circ \theta = \theta + \mu $$
We shall denote by $\Omega'^1 (M,S_k)$ the subspace of $\Omega^1 (M,S_k)$
constituted of sections of ${\cal{B}}_k$ under the correspondance (\ref{iso1}). 
Then $\omega = \theta + \mu$ obeys the equivariance and
horizontality conditions : 
\begin{eqnarray*}
    &(i)& R_{g}^* \omega = \bad{Ad} (g^{-1}) \omega , \; g \in GL^{k+1} \\
    &(ii)& \omega( \hat X_0 \oplus \cdots \oplus \hat X_k) = 
           X_0 \oplus \cdots \oplus X_k 
\end{eqnarray*}
and, for any $e_{k+2}$ above $e_{k+1}$, 
\begin{equation*}
    (iii) \,\,\,  X_{-1} \oplus \cdots \oplus X_k \mapsto \omega_{|e_{k+1}} 
          \left( e_{k+2} (X_{-1} \oplus \cdots \oplus X_k) \right) \,\,\, 
	  {\rm is \,\,\, inversible}
\end{equation*}
Reciprocally, if $\omega$ obeys $(i)$ and $(ii)$, then defining $\mu = \omega
- \theta$, we have $i_{\hat X} \mu = X - X = 0$ so horizontality, and
$R_{g}^*\mu = \bad{Ad} (g^{-1}) \mu$ by equivariance of $\theta$, so
$\mu \in \Omega^1 (M,S_k)$, with corresponding $\tilde{\mu}$. 
Next, as $\omega_{|e_{k+1}} 
          \left( e_{k+2} (X_{-1} \oplus \cdots \oplus X_k) \right) = X_{-1} +
	  \tilde{\mu}_{-1} X_{-1} \oplus \cdots \oplus X_k + \mu_k
	  X_{-1} $, $(iii)$ implies that in fact $1 + \tilde{\mu}$ is
inversible i.e. $\tilde{\mu} \in {\cal{B}}_k$. 
\\ \\
$\bullet$ 
For $\mu$, $\nu$ in $\Gamma({\cal B}_k)$, we can compose the isomorphisms 
$1+\tilde{\mu}$ and $1+\tilde{\nu}$ of $\got{gl}_{-1} \oplus \cdots \oplus 
\got{gl}_k$, at each $e_{k+1}$, to 
obtain $(1+\tilde{\mu}) (1+\tilde{\nu}) = 1 +\widetilde{\mu. \nu}$ 
(recall $B_k$ is a group). 
We have, from the equivariant point of vue  
$$ 
     \widetilde{\mu . \nu} = \tilde{\mu} + \tilde{\nu} + \tilde{\mu} \circ
     \tilde{\nu}_{-1}  
$$
and from the form point of vue   
$$
    \mu . \nu = \mu + \nu + i_{\nu} \mu
$$
where we see $\mu$ and $\nu$ as 1-forms valued in $\Gamma(TM_k)$ (section
\ref{sectionlocalspencercochains}) and $i_{\nu}$ is the
interior product extended to vector-valued forms. 
\\ \\
$\bullet$
Alternatively, we can see the bundle ${\cal{B}}_k$ as some jet space relative to
the differential operator $D_{\theta}$ previously defined. Indeed, defining at each
$e_{k+1}$, the equivalence relation : 
$$ 
    f_{k+1} \sim f'_{k+1} \,\,\, : \,\,\, 
    D_{\theta} f_{k+1} {_{|e_{k+1}}} = D_{\theta} f'_{k+1} {_{|e_{k+1}}} 
    \,\,\, , \,\,\, f_{k+1}, f_{k+1}' \in \rec{Aut}(M_{k+1})
$$
and denoting $[D_{\theta} f_{k+1}]_{|e_{k+1}}$ the resulting class, we build a bundle
associated to $M_{k+1}$ by considering the elements $[\tilde{D}_{\theta} f_{k+1}]$ with
equivariance under $GL^{k+1}$ inherited from the tensoriality of $D_{\theta} f_{k+1}$ : 
$$
     [\tilde{D}_{\theta} f_{k+1} ] \to \bad{Ad} (g^{-1}) \circ 
     [\tilde{D}_{\theta} f_{k+1}] \circ \bad{Ad} (g_0) 
$$
under $e_{k+1} \to e_{k+1}. g$.      
This allows us to identify this bundle with ${\cal{B}}_k$. 
\\ 
Now, for
$\tilde{\mu} \in {\cal{B}}_k$, written as 
$\tilde{\mu} = [\tilde{D}_{\theta} g_{k+1}]$ i.e. $\mu = [D_{\theta} g_{k+1}]$, 
the cocycle relation for $D_{\theta}$ passes to the jet equivalence to give : 
$$
    [D_{\theta}(g_{k+1} \circ f_{k+1})] = {f_{k+1}}^* [D_{\theta} g_{k+1}] +
    D_{\theta} f_{k+1}
$$
and induces the following action of $\rec{Aut} (M_{k+1})$ on $\Omega'^1
(M,S_k)$ : 
\begin{equation}
     \mu \to {f_{k+1}}^* \mu + D_{\theta} f_{k+1}
     \label{transformationdeformation} 
\end{equation}     
Next, for
$\tilde{\mu}, \tilde{\nu} \in {\cal{B}}_k$, written as $\tilde{\mu} =
[\tilde{D}_{\theta} f_{k+1}], \tilde{\nu} = [\tilde{D}_{\theta} g_{k+1}]$, that is 
$ \mu = [D_{\theta} f_{k+1}], \nu = [D_{\theta} g_{k+1}]$ , the same cocycle condition 
written from the equivariant point of vue (see 
\ref{actionofextdiffeoonlocalfields}) 
$$ 
    \tilde{D}_{\theta} (f_{k+1} \circ g_{k+1}) = 
    \tilde{D}_{\theta} f_{k+1} \circ 
    \left( 1 + \tilde{D}_{\theta} g_{k+1} \right) + \tilde{D}_{\theta} g 
$$
and conveniently rewritten as 
\begin{equation}
    1 + \tilde{D}_{\theta} (f_{k+1} \circ g_{k+1}) = 
    \left( 1 + \tilde{D}_{\theta} f_{k+1} \right) 
    \left( 1 + \tilde{D}_{\theta} g_{k+1} \right)
    \label{compojetdef} 
\end{equation}
where $1 + \tilde{D}_{\theta} f_{k+1}$ is evaluated at the point $g_{k+1}
(e_{k+1})$ and $1 + \tilde{D}_{\theta} g_{k+1}$ at $e_{k+1}$ as stated in
(\ref{actionextendeddiffeo}), 
passes to the jet equivalence, and give us back the composition of deformations
:
$$  1 + \widetilde{\mu.\nu} = 
    1 + 
    [\tilde{D}_{\theta} (f_{k+1} \circ g_{k+1})] = 
    \left( 1 + [\tilde{D}_{\theta} f_{k+1}] \right) 
    \left( 1 + [\tilde{D}_{\theta} g_{k+1}] \right)
    = 
    ( 1 +\tilde{\mu} ) ( 1 +\tilde{\nu} )  
$$      
i.e. the composition in $\rec{Aut}(M_{k+1})$ induces at the jet level the
composition of deformations. 

\subsubsection{Deformed frame bundle}

Now, we analyse the deformations from another point of vue, perhaps more
concrete, and we show how to rederive in this context the results given above,
and how it allows to produce new ones. \\ \\ 
$\bullet$ 
For $\tilde{\mu} \in \Gamma({\cal{B}}_k)$, we notice that, for $l \leq k$,
the section $\tilde{\mu}_{-1} \oplus \cdots \oplus \tilde{\mu}_l$ is
equivariant under $GL^{l+1}$ and invariant under $GL_{l+2} \ltimes \cdots
\ltimes GL_k$, so we can descend 
$\tilde{\mu}_{-1} \oplus \cdots \oplus \tilde{\mu}_l$ to a section of
$\Gamma ({\cal{B}}_l)$ that is we can view it as a equivariant function on
$M_{l+1}$. 
\\ \\
$\bullet$ 
We define $M_{0,\mu}$ as the space of $e_{0,\mu}$'s obtained as 
$$ e_{0,\mu} = e_0 \circ (1 + \tilde{\mu}_{-1})_{|e_0}^{-1} $$
Thus $M_{0,\mu}$ is a $GL_0$ principal bundle over $M$ which is in fact,
here, $M_0$ (as here $1+\tilde{\mu}_{-1}$ is a gauge transformation). We
denote : 
$$
    F_{-1,\mu} : M_0 \to M_{0,\mu} , \; 
                   e_0 \mapsto e_0 \circ (1 + \tilde{\mu}_{-1})_{|e_0}^{-1} 
$$
This map is a principal bundle isomorphism inducing the identity on the base $M$. 
\\ \\
$\bullet$ 
Next, define $M_{1,\mu}$ as the space of $e_{1,\mu}$'s obtained as : 
$$
    e_{1,\mu} = F_{-1,\mu}{_*} e_1 \circ (1 + \tilde{\mu}_{-1} 
    \oplus \tilde{\mu}_0 )_{|e_1}^{-1} 
$$
These are linear frame above $M_{0,\mu}$ since 
\begin{eqnarray*}
    &(i)& \pi_{0,-1}{_*} e_{1,\mu} (X_{-1} \oplus X_0) 
           = e_{0,\mu} (X_{-1}) \\    		    
    &(ii)& e_{1,\mu} (X_0) = \hat X_0
\end{eqnarray*}
where this comes from the fact $F_{-1,\mu}$ is a principal bundle
isomorphism.  
Again, $M_{1,\mu}$ is a $GL_1$-principal bundle above $M_{0,\mu}$ and a
$GL^1$-principal bundle above $M$. We define 
$$
    F_{0,\mu} : M_1 \to M_{1,\mu} , \; 
                   e_1 \mapsto F_{-1,\mu}{_*} 
		   e_1 \circ (1 + \tilde{\mu}_{-1}  \oplus 
		       \tilde{\mu}_0 )_{|e_1}^{-1}
$$
This map is a principal bundle isomorphism. 
\\ \\
$\bullet$ 
Recursively, we define $M_{l+1,\mu}$ as the space of $e_{l+1,\mu}$'s
obtained as 
\begin{equation}
    e_{l+1,\mu} = F_{l-1,\mu} {_*} e_{l+1} \circ 
     (1 + \tilde{\mu}_{-1}  \oplus \cdots \oplus 
          \tilde{\mu}_l )_{|e_{l+1}}^{-1} 
    \label{deformedframe}
\end{equation}	  	
Using the fact that the precedingly constructed $F_{l-1,\mu}$ is a principal
bundle isomorphism, we show that $e_{l+1,\mu}$ are linear frames above
$M_{l,\mu}$, and obtain the principal bundles 
\begin{equation}
     \xymatrix{
     GL_{l+1} \ar[r] & M_{l+1,\mu} \ar[r]^-{\pi'_{l+1,l}} & M_{l,\mu} 
     }
\end{equation}
\begin{equation}
     \xymatrix{ 
     GL^{l+1} \ar[r] & M_{l+1,\mu} \ar[r]^-{\pi'_{l+1,-1}} & M 
     }    
\end{equation}
In summary, we have interpreted $\tilde{\mu}$ as providing an iterative
fibering encoded in the commutative diagram : 
\begin{equation*}
     \xymatrix{ 
     M_{k+1} \ar[r] \ar[d]^-{F_{k,\mu}} & 
     M_k \ar[r] \ar[d]^-{F_{k-1,\mu}} & 
     \cdots \ar[r] & 
     M_1 \ar[r] \ar[d]^-{F_{0,\mu}} & 
     M_0 \ar[r] \ar[d]^-{F_{-1,\mu}} &
     M_{-1} \\      
     M_{k+1,\mu} \ar[r] & 
     M_{k,\mu} \ar[r] & 
     \cdots \ar[r] & 
     M_{1,\mu} \ar[r] & 
     M_{0,\mu} \ar[r] & 
     M_{-1} 
     }
\end{equation*}     
that is $\pi'_{l+1,l} \circ F_{l+1,\mu} = F_{l,\mu} \circ \pi_{l+1,l}$,
with commutation of the subsquares (covariance of $F_{l-1,\mu}$) : 
\begin{equation}
    \xymatrix{
    GL_{l'+1} \ltimes \cdots \ltimes GL_l \ar[r] \ar@{=}[d] & 
    M_{l} \ar[r] \ar[d]^-{F_{l-1,\mu}} & 
    M_{l'} \ar[d]^-{F_{l'-1,\mu}} \\ 
    GL_{l'+1} \ltimes \cdots \ltimes GL_l \ar[r] & 
    M_{l,\mu} \ar[r] & 
    M_{l',\mu} 
    } 
\end{equation}  
Note that if a deformation $\mu$ is a $\partial$-cocycle, i.e. $\partial
\tilde{\mu}_{|e_{k+1}} = 0$ at each $e_{k+1}$, then the induced deformation is
simply a gauge transformation $g^{k+1} \in {\cal{GL}}^{k+1}$ of $M_{k+1}$ 
whose equivariant form $\tilde{g} = \tilde{g}^{k+1}$ satisfies (see section
\ref{sectionthejetaction}) : 
$ \bad{Ad} (\tilde{g}^{-1}) = 1 + \tilde{\mu} $. 
\subsubsection{Deformed frame form}

As each $M_{l+1,\mu}$ is a bundle of $(l+2)$-linear frames above
$M_{l,\mu}$, we can dually define the frame form. On $M_{k+1,\mu}$, define
the deformed frame form at $e_{k+1,\mu}$ by  
\begin{equation}
     \theta_{\mu}^k = {e_{k+1,\mu}}^{-1} \pi'_{k+1,k}{_*} 
\end{equation}
Then, by construction of the $e_{l,\mu}$'s, $\theta_{\mu}^k$ satisfies the
same properties of equivariance, horizontality, and recursion as the ordinary
frame form on $M_{k+1}$. Moreover we have from (\ref{deformedframe}) 
\begin{eqnarray*}
    \theta_{\mu}^k 
    &=& 
    {e_{k+1,\mu}}^{-1} \pi'_{k+1,k}{_*} \\
    &=& 
    ( 1 +\tilde{\mu}_{-1} \oplus \cdots \tilde{\mu}_k )_{|e_{k+1}}
    {e_{k+1}}^{-1} {F_{k-1,\mu}}^{-1} {_*} \pi'_{k+1,k}{_*} \\
    &=& 
    (1 + \tilde{\mu})_{|e_{k+1}} {e_{k+1}}^{-1} \pi_{k+1,k}{_*}
    {F_{k,\mu}}^{-1} {_*} 
\end{eqnarray*}
that is the deformed frame form is related to the frame from on $M_{k+1}$
thanks to 
\begin{equation}
     {F_{k,\mu}}^* \theta_{\mu}^k = (1 + \tilde{\mu} ) \circ \theta^k 
      = \theta + \mu
     \label{relationframeforms}  
\end{equation}                 
The deformed curvature is defined as 
\begin{equation*}
     \Theta_{\mu}^{k-1} = d \theta_{\mu}^k + \frac{1}{2} 
      [ \theta_{\mu}^k , \theta_{\mu}^k ] \mod \got{h}_{k-1}
\end{equation*}
and is null iff the frames $e_{k+1,\mu}$ are indeed jet frames (this being a
consequence of section \ref{sectionlinearframes}). Next, computing the deformed curvature from
(\ref{relationframeforms}), we have : 
\begin{equation*}
    {F_{k,\mu}}^* \Theta_{\mu}^{k-1} = \rec{d} ( \theta + \mu ) 
     + \frac{1}{2} [ \theta + \mu , \theta + \mu ] \mod \got{h}_{k-1} 
\end{equation*}
With all this in mind, we define the second Spencer operator as 
\begin{equation}
    D_{\theta} \mu = d ( \theta + \mu ) + \frac{1}{2} [ \theta + \mu ,
    \theta + \mu ] \mod \got{h}_{k-1} 
\end{equation}
Then, without anymore calculations, it becomes clear from the deformed frames point of vue
that $D_{\theta} \mu$ is a $\got{gl}_{-1} \oplus \cdots \oplus \got{gl}_{k-1}$-valued 
tensorial 2-form on $M_{k+1}$, which is null iff the deformed frame bundle
$M_{k+1,\mu}$ is actually the jet frame bundle $M_{k+1}$. 

\subsubsection{Extended diffeomorphisms action}
         
We shall now derive, from the deformed bundle point of vue, the transformation
of $\tilde{\mu}$ and $\mu$ under $\rec{Aut}(M_{k+1})$, that is, we explain
where does come from the transformation law $\mu \to {f_{k+1}}^* \mu + D
f_{k+1}$, equation (\ref{transformationdeformation}).  
\\ \\
Take $f_{k+1} \in \rec{Aut}(M_{k+1})$, denote by $f_{l}$ its
projections on $\rec{Aut}(M_l)$, and call $\tilde{\mu}'$ the transformed of 
$\tilde{\mu}$. \\ \\
$\bullet$ 
To first order, we define $\tilde{\mu}'$ uniquely from : 
\begin{equation*}
    f_0(e_0) \circ ( 1 + \tilde{\mu}_{-1})_{|f_0(e_0)}^{-1} 
    = f_{-1}{_*} e_0 \circ (1 + \tilde{\mu}'_{-1} )_{|e_0}^{-1} 
\end{equation*}
Thus, we have : 
\begin{equation*}
    (1 + \tilde{\mu}'_{-1})_{|e_0} = ( 1 + {f_0}^* \tilde{\mu}_{-1} 
    )_{|e_0} \circ ( (f_0 (e_0))^{-1} \circ f_{-1}{_*} e_0 )
\end{equation*}
Now, from the section point of vue, $D_{\theta} f_0$ is such that (see section
\ref{actionofextdiffeoonlocalfields}) : 
\begin{equation}
     1 + \tilde{D}_{\theta} f_0 = (f_0 (e_0))^{-1} \circ f_{-1}{_*} e_0 
\end{equation}
so that we find 
\begin{equation}
    1 + \tilde{\mu}'_{-1} = ( 1 + {f_0}^* \tilde{\mu}_{-1} ) \circ 
     (1 + \tilde{D}_{\theta} f_0 ) 
\end{equation}
In one word, we have constructed the commutative square 
\begin{equation*}
     \xymatrix{ 
     M_0 \ar[r]^-{f_0}  \ar[d]^-{F_{-1,\mu'}} & 
     M_0 \ar[d]^-{F_{-1,\mu}} \\
     M_{0,\mu'} \ar[r]^-{f_{-1}{_*}} & 
     M_{0,\mu}
     }
\end{equation*}
since we have $ F_{-1,\mu} \circ f_0 {_{|e_0}} = f_0{_{|e_0}} \circ (1 +
\tilde{\mu}_{-1})_{|f_0(e_0)} $. We define the intertwining diffeomorphism 
$$ 
    f_{0,\mu} = F_{-1,\mu} \circ f_0 \circ {F_{-1,\mu'}}^{-1}
$$
as a useful object for later purpose. 
\\ \\
$\bullet$ 
To second order, we define in the same way $\tilde{\mu}'$ from the
commutative square (note the appearance of the intertwining diffeomorphism at this
level)
\begin{equation*}
     \xymatrix{
     M_1 \ar[r]^-{f_1} \ar[d]^-{F_{0,\mu'}} & 
     M_1 \ar[d]^-{F_{0,\mu}} \\
     M_{1,\mu'} \ar[r]^-{f_{0,\mu}{_*}} & M_{1,\mu} 
     }           
\end{equation*}
that is :
$$
    F_{-1,\mu}{_*} f_1(e_1) \circ  (1 + \tilde{\mu}_{-1} \oplus
    \tilde{\mu}_0)_{|f_1(e_1)}^{-1} 
    = 
    f_{0,\mu}{_*} F_{-1,\mu'}{_*} e_1 \circ 
    ( 1 + \tilde{\mu}'_{-1}\oplus \tilde{\mu}'_0 )_{|e_1}^{-1}
$$
This is equivalent, from the definition of $f_{0,\mu}$, to 
$$
    f_1(e_1) \circ  (1 + \tilde{\mu}_{-1} \oplus
    \tilde{\mu}_0 )_{|f_1(e_1)}^{-1}  
    =                            
    f_0{_*} e_1 \circ 
    ( 1 + \tilde{\mu}'_{-1} \oplus \tilde{\mu}'_0 )_{|e_1}^{-1}
$$
and, by the same reasoning as for the first order case, this proves 
\begin{equation*}
    (1 + \tilde{\mu}'_{-1} \oplus \tilde{\mu}'_0)  = 
    ( 1 + {f_1}^*\tilde{\mu}_{-1} \oplus {f_1}^* \tilde{\mu}_0 ) \circ
    ( 1 + \tilde{D}_{\theta} f_1 ) 
\end{equation*}
Note that this is consistent with the first order result since this last
equation implies, by
invariance of $\mu_{-1}$ with respect to $GL_1$ and graded action of $1 +
\tilde{D}_{\theta} f_1$, $1 + \tilde{\mu}'_{-1} = 
(1 + {f_0}^*\tilde{\mu}_{-1}) \circ (1 + \tilde{D}_{\theta} f_0) $. 
\\ \\
$\bullet$ 
Recursively, if we have defined the action at the $M_l$ level, obtaining the
commutative square 
\begin{equation*}
    \xymatrix{
    M_l \ar[r]^-{f_l} \ar[d]^-{F_{l-1,\mu'}} &
    M_l \ar[d]^-{F_{l-1,\mu}} \\
    M_{l,\mu'} \ar[r]^-{f_{l-1,\mu}{_*}} & 
    M_{l,\mu} 
    }
\end{equation*}    
we define the intertwining diffeomorphism $f_{l,\mu} = F_{l-1,\mu} \circ f_l \circ
{F_{l-1,\mu'}}^{-1}$, and $\tilde{\mu}'$ by the commutative square at next
level 
\begin{equation*}
    \xymatrix{
    M_{l+1} \ar[r]^-{f_{l+1}} \ar[d]^-{F_{l,\mu'}} &
    M_{l+1} \ar[d]^-{F_{l,\mu}} \\
    M_{l+1,\mu'} \ar[r]^-{f_{l,\mu}{_*}} & 
    M_{l+1,\mu} 
    }
\end{equation*}    
This means $F_{l,\mu} (f_{l+1}(e_{l+1})) = f_{l,\mu}{_*}
F_{l,\mu'}(e_{l+1}) $, that is : 
\begin{eqnarray*}
    &F_{l-1,\mu}{_*} f_{l+1}(e_{l+1}) \circ ( 1 + \tilde{\mu}_{-1}
     \oplus \cdots \oplus \tilde{\mu}_l  
    )_{|f_{l+1}(e_{l+1})}^{-1} 
    = \qquad \qquad \qquad \qquad \qquad& \\
    & \qquad \qquad \qquad \qquad \qquad
    f_{l,\mu}{_*} F_{l-1,\mu'}{_*} e_{l+1} \circ ( 1 + 
    \tilde{\mu}'_{-1} \oplus \cdots \oplus 
    \tilde{\mu}'_l  )_{|e_{l+1}}^{-1} & 
\end{eqnarray*}
and, thanks to the definition of $f_{l,\mu}$, 
$$
    f_{l+1}(e_{l+1}) \circ ( 1 + 
    \tilde{\mu}_{-1}  \oplus \cdots \oplus 
    \tilde{\mu}_l  )_{|f_{l+1}(e_{l+1})}^{-1} 
    = 
    f_l {_*} e_{l+1} \circ ( 1 + 
    \tilde{\mu}'_{-1}  \oplus \cdots \oplus \tilde{\mu}'_l
    )_{|e_{l+1}}^{-1}
$$
Now, using the fact 
$$
    ( 1 + \tilde{D}_{\theta} f_{l+1} )_{|e_{l+1}} = (f_{l+1}(e_{l+1}))^{-1} \circ f_l{_*}
    e_{l+1} 
$$
we obtain the transformation law 
$$
    (1 + \tilde{\mu}'_{-1} \oplus \cdots \oplus \tilde{\mu}'_l )_{|e_{l+1}} = 
    ( 1 + \tilde{\mu}_{-1} \oplus \cdots \oplus \tilde{\mu}_l
    )_{|f_{l+1}(e_{l+1})} \circ ( 1 + \tilde{D}_{\theta} f_{l+1})_{|e_{l+1}} 
$$
Finally, we have obtained the action of $f_{k+1} \in \rec{Aut}(M_{k+1})$ on
$\tilde{\mu}$ in the form : 
\begin{equation}
    1+\tilde{\mu} \to 1+\tilde{\mu}' = 
    (1 + {f_{k+1}}^* \tilde{\mu} ) \circ ( 1 +\tilde{D}_{\theta} f_{k+1}) 
    \label{actiondiffsection}
\end{equation}    
In the form language, from section \ref{actionofextdiffeoonlocalfields}, 
the equation (\ref{actiondiffsection}) becomes 
\begin{equation*}
     \theta + \mu' = {f_{k+1}}^* ( \theta + \mu )
\end{equation*}
and we recover the transformation law 
\begin{equation}
     \mu \to \mu' = {f_{k+1}}^* \mu + D_{\theta} f_{k+1} 
\end{equation}
Note that the intertwining diffeomeorphisms $f_{k,\mu}$ not only depend on the 
transformation $f_{k+1}$ but also on the deformation $\mu$. Infinitesimally,
this 
difference between $f_{k+1}$ and $f_{k,\mu}$ is reflected, at least in 2D CFT,
by 'field dependant ghosts' \cite{serge} originally introduced in \cite{becchi}.       
\subsubsection{Action of deformations on local fields}

We now look for the action of deformations on local fields, in the same way as
in section \ref{actionofextdiffeoonlocalfields}. For a deformation $\mu$ of
$M_{k+1}$, as the deformed frame bundle $M_{k+1,\mu}$ is also principal, we can
speak of the local fields on $M_{k+1,\mu}$, by doing the same construction as in
section \ref{sectionlocalspencercochains}, with $M_{k+1}$ replaced by
$M_{k+1,\mu}$. We shall denote $S_{k,\mu}$ the deformed bundle 
$$ 
S'_{k,\mu} = M_{k+1,\mu} \times_{\bad{Ad}} \got{gl}_{-1} \oplus \cdots \oplus
\got{gl}_k 
$$
Take a local field $\alpha' \in \Omega^* (M,S_{k,\mu})$. 
\\ \\
$\bullet$ From the tensorial form point of view, the action of a deformation is
to read the form $\alpha'$ on $M_{k+1}$ by pullback i.e. : 
\begin{equation}
     \alpha' \to \alpha = {F_{k,\mu}}^* \alpha'
     \label{actiondeformation}
\end{equation}
This is consistent since $F_{k,\mu}$ is a principal bundle isomorphism. 
\\ \\
$\bullet$ Viewing $\alpha'$ as a equivariant function $\tilde{\alpha}'$ on
$M_{k+1,\mu}$, thanks to the formula 
$$ 
     \alpha' = \tilde{\alpha}' \circ \theta_{\mu}
$$
the transformation (\ref{actiondeformation}) now reads : 
\begin{equation}
     \tilde{\alpha}' \to \tilde{\alpha} = {F_{k,\mu}}^* \tilde{\alpha} \circ (1
     + \tilde{\mu} )
\end{equation}
This is obtained with a calculation similar to that establishing 
(\ref{actionextendeddiffeo}).  
\subsubsection{Synthesis}
\label{synthesisdeformation}

$\bullet$ 
We have obtained thus the operator $D_{\theta}: \Omega'^1 (M,S_k) \to
\Omega^2(M,S_{k-1})$, acting on deformations as 
\begin{equation*}
    D_{\theta} \mu 
    = d (\theta + \mu) + \frac{1}{2} [\theta + \mu , \theta +
           \mu] \mod \got{h}_{k-1}     	   
\end{equation*}
We can alternatively write, using the structure equation $\Theta^{k-1} = 0$, 
\begin{equation}
     D_{\theta} \mu = d_{\theta} \mu + \frac{1}{2} [ \mu , \mu ] 
     \mod \got{h}_{k-1} 
     \label{secondDdefinition}
\end{equation}
This is the definition of $D_{\theta}$ we will take. 
\\ 
More generally, for $\omega$ a $\got{gl}_{-1} \oplus \cdots \oplus \got{gl}_k$-valued
1-form on $M_{k+1}$ satisfying properties $(i),(ii),(iii)$ of a deformation (see
section \ref{sectiondeformationspace}), we
define 
$$ D_{\omega} \mu = d_{\omega} \mu + \frac{1}{2} [\mu,\mu] \mod \got{h}_{k-1}
$$
and, for technical purpose 
$$
   {\underline{D}}_{\omega} \mu = d (\omega + \mu) + \frac{1}{2} [ \omega + \mu ,
   \omega + \mu] \mod \got{h}_{k-1}
$$   
Note that we then have 
\begin{equation}
     {\underline{D}}_{\omega} \mu   
     = 
     d_{\omega} \omega + D_{\omega} \mu = 
     d \omega + \frac{1}{2} [ \omega, \omega ] + D_{\omega} \mu \mod
     \got{h}_{k-1}
     \label{technicalcurvature}
\end{equation}
Then $D_{\omega} \mu $ and ${\underline{D}}_{\omega} \mu$ are still tensorial 
i.e. in $\Omega^2(M,S_{k-1})$. We
shall call the quantities $D_{\omega} \mu$ and 
${\underline{D}}_{\omega} \mu$ torsion or curvature, as these concepts are not to
be distinguished in Cartan geometry. These definitions can also be used on any of the
deformed frame bundles. \\       
As for the symmetries, the properties of $D_{\theta}$ are summarised in : 
\\ \\
{\it $D_{\theta}$ defines a cocycle on the space of deformations $\Gamma ({\cal{B}}_k)
\simeq \Omega'^1(M,S_k)$,
seen as a group, with values in $\Omega^2 (M,S_{k-1})$ (see eq. (\ref{defcocycle}) 
hereafter for the explicit cocycle law). Its kernel contains the
deformations induced by $\rec{Aut} (M_{k+1})$, that is $D_{\theta} \mu = 0$, for
$\mu = D_{\theta} f_{k+1}$. 
} 
\\ \\ 
$\bullet$ 
We prove first the cocycle property. 
To a deformation $\mu$ seen as an equivariant function $\tilde{\mu}$, we
associate the deformation $\tilde{\mu}' = \tilde{\mu} \circ
{F_{k,\nu}}^{-1}$ on the deformed frame bundle $M_{k+1,\nu}$, and so the
corresponding $\mu' = \tilde{\mu} \circ \theta_{\nu}$. 
\\
Then, from the point of vue of $M_{k+1,\nu}$, the curvature is obtained as : 
\begin{equation}
     {\underline{D}}_{\theta_{\nu}} \mu' = d (\theta_{\nu} + \mu') + 
     \frac{1}{2} [\theta_{\nu} + \mu' , \theta_{\nu} + \mu' ] 
     \mod \got{h}_{k-1} 
\end{equation} 
Next, this curvature form is read on $M_{k+1}$ via the pullback ${F_{k,\nu}}^*
{\underline{D}}_{\theta_{\nu}} \mu'$. We have : 
\begin{eqnarray*}
    {F_{k,\nu}}^* ({\underline{D}}_{\theta_{\nu}} \mu') 
    & = & 
    {F_{k,\nu}}^* \left( d ( \theta_{\nu} + \mu' ) 
                         + \frac{1}{2} [ \theta_{\nu} + \mu' ,
			 \theta_{\nu} + \mu' ] \right) \mod \got{h}_{k-1}
    \\
    & = & 
    d (\theta + \nu + {F_{k,\nu}}^* \mu') + 
    \frac{1}{2} [ \theta + \nu + {F_{k,\nu}}^* \mu' , 
                  \theta + \nu + {F_{k,\nu}}^* \mu' ] \mod \got{h}_{k-1}
\end{eqnarray*}		  			 
Now, ${F_{k,\nu}}^* \mu'$ is the deformation $\mu$ deformed by $\nu$,
since (compare with equation (\ref{actiondeformation})) 
\begin{eqnarray*}
    {F_{k,\nu}}^* \mu' 
    & = & 
    \mu' \circ {F_{k,\nu}}{_*} 
    \\ 
    & = & 
    \tilde{\mu}' \circ \theta_{\nu} \circ {F_{k,\nu}}{_*} \\
    & = & 
    \tilde{\mu} \circ {F_{k,\nu}}^* \theta_{\nu} \\
    & = &
    \tilde{\mu} \circ (\theta + \nu ) \\
    & = & 
    \mu + i_{\nu} \mu	
\end{eqnarray*}    		 
So, we obtain : 
\begin{equation*}
     {\underline{D}}_{\theta} ( \mu . \nu ) = 
     {F_{k,\nu}}^* {\underline{D}}_{\theta_{\nu}} \mu' 
\end{equation*}
Then, this last equation can be rewritten thanks to (\ref{technicalcurvature})
as a cocycle law for $D_{\theta}$ (recall the action of deformations 
(\ref{actiondeformation})) : 
\begin{equation}
    D_{\theta} ( \mu.\nu  ) = {F_{k,\nu}}^* D_{\theta_{\nu}} \mu' + D_{\theta}
    \nu
    \label{defcocycle}
\end{equation}
$\bullet$ Now, we prove the nilpotency. 
For $\mu = D_{\theta} f_{k+1}$, we have thanks to the structure equation 
\begin{eqnarray*}    
    D_{\theta} D_{\theta} f_{k+1} & = & d ( {f_{k+1}}^* \theta^k ) + 
       \frac{1}{2} [{f_{k+1}}^* \theta^k , 
      {f_{k+1}}^* \theta^k ] \mod \got{h}_{k-1} \\
      & = & {f_{k+1}}^* \Theta^{k-1} \\ 
      & = & 0 
\end{eqnarray*}
\begin{flushright} $\blacksquare$ \end{flushright}
$\bullet$ All this is summarised in the sequence : 
\begin{equation}
     \xymatrix{ 
     \rec{Aut} (M_{k+1} ) \ar[r]^-{D_{\theta}} & 
     \Omega'^1 (M,S_k) \ar[r]^-{D_{\theta}} & 
     \Omega^2 (M,S_{k-1}) 
     }
     \label{spencer12}
\end{equation}
We have, as stated in \cite{pom} : 
\\ \\ 
{\it The non linear complex (\ref{spencer12}) is locally exact i.e. on a suitable
open cover $(U_i)$ of $M$, the equation $D_{\theta} \mu = 0$ on $U_i$ implies 
$$
    \mu = D_{\theta} f_{k+1,i} \,\,\, {\rm{for}} \,\,\, f_{k+1,i} \in
    \rec{Aut}(U_{i,k+1})
$$
}
\\
A proof of this in local coordinate form is given in \cite{pom}. 
Here, we shall indicate another way to see this, using Cartan geometry
\cite{sharpe}.     
We work on a chart
$(U_i,\varphi_i)$ of $M$, with (invertible) maps $\varphi_i : U_i \to \mathbb{R}^n$, and $U_i$
contractible. Thanks to the 'fundamental theorem of calculus' of \cite{sharpe}, 
the condition $D_{\theta} \mu =
0$, written $d \omega + \frac{1}{2} [ \omega , \omega ] = 0 \mod \got{h}_{k-1}$,
, $\omega = \theta + \mu$, proves that there exists locally on $U_i$, a map 
$$ \phi_{k+1,i} : U_{i,k+1} \to \mathbb{R}^n{_{k+1}} $$
such that 
$$ \omega = \theta + \mu = {\phi_{k+1,i}}^* \tilde{\theta} $$
where $\tilde{\theta}$ is the frame form on $\mathbb{R}^n{_{k+1}}$. Then,
equivariance of $\theta + \mu$ and $\tilde{\theta}$, and evaluation on
frames, proves that $\phi_{k+1,i}$ is indeed a principal bundle isomorphism,
locally defined above $U_i$. 
\\ \\
Moreover, the prolongation $\varphi_{k+1,i} : U_{i,k+1}
\to \mathbb{R}^n{_{k+1}}$, which satisfies by construction ${\varphi_{k+1,i}}^*
\tilde{\theta}  = \theta$, enables us to define 
$$
     f_{k+1,i}  = {\varphi_{k+1,i}}^{-1} \circ \phi_{k+1,i}
$$
such that $f_{k+1,i} \in \rec{Aut} (M_{k+1})$. In this way, we obtain : 
\begin{eqnarray*}
     \mu & = & {\phi_{k+1,i}}^* \tilde{\theta} - \theta \\ 
            & = & (\varphi_{k+1,i} \circ f_{k+1,i})^{*} \tilde{\theta} - \theta
	    \\
	    & = & {f_{k+1,i}}^* {\varphi_{k+1,i}}^* \tilde{\theta} - \theta \\
	    & = & {f_{k+1,i}}^{*} \theta - \theta \\
	    & = & D_{\theta} f_{k+1,i} 
\end{eqnarray*}
This means the sequence (\ref{spencer12}) is locally exact at
$\Omega'^1(M,S_k)$. This construction can be summarised in the 
commutative square where each arrow is a principal bundle morphism : 
\begin{equation*}
    \xymatrix{ 
    U_{i,k+1} \ar[r]^-{\phi_{k+1,i}} \ar[d]^-{f_{k+1,i}} & 
    {\mathbb{R}^n}_{k+1} \ar@{=}[d] \\
    U_{i,k+1} \ar[r]^-{\varphi_{k+1,i}} & 
    {\mathbb{R}^n}_{k+1} 
    }
\end{equation*}    
The map $\phi_{k+1,i}$ is a development map \cite{sharpe}, here adapted to the diffeomorphism
symmetry. 
\begin{flushright} $\blacksquare$ \end{flushright} 
           
\subsection{Synthesis}

\subsubsection{Symmetries and deformations}

$\bullet$ 
The study of symmetries and deformations in the language of linear frames
reveals that they have the same structure, as shown in the covariant and
commutative diagrams : 
\begin{equation*}
    \xymatrix{ 
    M_{k+1} \ar[r] \ar[d]^-{f_{k+1}} & 
    M_k \ar[r] \ar[d]^-{f_k} & 
    \cdots \ar[r] &
    M_1 \ar[r] \ar[d]^-{f_1} & 
    M_0 \ar[r] \ar[d]^-{f_0} & 
    M_{-1} \ar[d]^-{f_{-1}} \\
    M_{k+1} \ar[r] & 
    M_k \ar[r] & 
    \cdots \ar[r] & 
    M_1 \ar[r] &
    M_0 \ar[r] & 
    M_{-1}
    }
\end{equation*}    
for the symmetries, and similarly 
\begin{equation*}
     \xymatrix{ 
     M_{k+1} \ar[r] \ar[d]^-{F_{k,\mu}} & 
     M_k \ar[r] \ar[d]^-{F_{k-1,\mu}} & 
     \cdots \ar[r] & 
     M_1 \ar[r] \ar[d]^-{F_{0,\mu}} & 
     M_0 \ar[r] \ar[d]^-{F_{-1,\mu}} &
     M_{-1} \ar[d]^-{\rec{id}} \\      
     M_{k+1,\mu} \ar[r] & 
     M_{k,\mu} \ar[r] & 
     \cdots \ar[r] & 
     M_{1,\mu} \ar[r] & 
     M_{0,\mu} \ar[r] & 
     M_{-1} 
     }
\end{equation*}
for the deformations. From a gravity point of vue, the similarity between these
two structures is natural as one can understand them in term of a 
generalised equivalence principle : the {\it{gravitational fields}} $\mu$ of the
second diagram are
'locally' equivalent, i.e. in fact at the level of jets (see section
\ref{sectiondeformationspace}), to the {\it{general changes of coordinate frame}}
$f_{k+1}$ of the first diagram (see e.g. \cite{newman} for the use of Cartan
geometry in gravity). Alternatively, one can also think of the deformations
$\mu$ as generalised Beltrami differentials \cite{serge}, the equation  
$\mu = D_{\theta} f_{k+1}$ being then a generalised Beltrami equation, with
integrability conditions $D_{\theta} \mu = 0$. This fact will be
further studied elsewhere. The interesting fact here is that both symmetries,
i.e. $\rec{Aut}(M_{k+1})$, and fields, i.e. deformations $\Omega'^1(M,S_k)$,
appear on the same footing.     
\\ \\
$\bullet$ 
Alternatively, as $\rec{Aut}(M_{k+1})$ acts on
$\Omega'^1(M,S_k) \simeq \Gamma ({\cal{B}}_k)$, and as $\Omega'^1(M,S_k)$ is a group, we can consider the
group semi-direct product 
\begin{equation}
 \Omega'^1 (M,S_k) \rtimes \rec{Aut}(M_{k+1}) 
 \label{productspace}
\end{equation}  
as encoding the preceding two diagrams in a unified manner. The group law is
explicitly given by 
\begin{equation}
     (\mu , f_{k+1} ) .  (\mu' , f'_{k+1}) = 
     ( \mu. (f_{k+1}.\mu') , f'_{k+1} \circ f_{k+1} ) 
     \label{groupactionperrot}
\end{equation}
In this equation, $f.\mu' = {f_{k+1}}^* \mu' + D_{\theta} f_{k+1}$ is the (right) action
of $f_{k+1}$ on $\mu'$, and $\mu. \nu$ with $\nu = f_{k+1}.\mu'$ denotes the
composition of deformations. We have $f'_{k+1} \circ f_{k+1}$ on the r.h.s. because of
pull-back law. 
\\ \\
This structure is roughly speaking some non linear analogue to
the one in \cite{perrot} used for treating diffeomorphisms. Maybe one could use
this to derive, as in \cite{perrot}, some cohomological structure related to the
BRS one. In this respect, as it is natural to view the space $\Omega'^1(M,S_k)$
as a classifying space for $\rec{Aut}(M_{k+1})$ by analogy with gauge theory, we
can also view the product (\ref{productspace}) as giving rise to the
equivariant cohomology type quotient : 
\begin{equation*}
     \xymatrix{ 
     \rec{Aut}(M_{k+1}) \ar[r] & 
     \rec{Aut}(M_{k+1}) \times \Omega'^1(M,S_k) \ar[r] & 
     \rec{Aut}(M_{k+1}) \times_{\rec{Aut}(M_{k+1})} \Omega'^1 (M,S_k) 
     }
\end{equation*}
where $\rec{Aut}(M_{k+1})$ acts on both sides of the product as in 
(\ref{groupactionperrot}) with $\mu = 0$. 
      
\subsubsection{Non linear Spencer sequences}

$\bullet$
The two sequences (\ref{spencer11}) and (\ref{spencer12}) enable us to
construct the non linear Spencer sequence of \cite{pom} as : 
\begin{equation}
\xymatrix{
    \rec{id} \ar[r] & 
    \rec{Aut}(M) \ar[r]^-{j_{k+1}} & 
    \rec{Aut} (M_{k+1}) \ar[r]^-{D_{\theta}} & 
    \Omega'^1(M,S_k) \ar[r]^-{D_{\theta}} & 
    \Omega^2(M,S_{k-1}) \ar[r] & 
    0 
}
\label{firstspencersequence}
\end{equation}
This sequence is then globally exact at $\rec{Aut}(M)$ and $\rec{Aut}(M_{k+1})$,
and locally exact at
$\Omega'^1(M,S_k)$. This sequence embodies all the structure necessary for
gravity theories : from left to right, we have the base space symmetry, then
the frame space symmetry, then the gravity potentials (deformations), and
finally the gravity field strenghts (curvatures).
\\
For any deformation $\mu$, we also have Bianchi type identities in the form 
\begin{eqnarray*}
    d_{\theta + \mu} D_{\theta} \mu 
    & = & 
    d_{\theta + \mu} d_{\theta + \mu} (\theta + \mu) \\
    & = & 0 
\end{eqnarray*}
This fact indicates that if we want to prolongate the non linear Spencer sequence 
(\ref{firstspencersequence}) we have to intertwine the differential operators 
involved with $\mu$ fields, such as 
$d_{\theta + \mu} : \Omega^2 (M,S_{k-1}) \to
\Omega^3 (M,S_{k-2})$ here. This means one cannot extend the non linear Spencer
sequence to forms of degree $>2$ without introducing more fields, in analogy
with the fact that one cannot extend non abelian \u{C}ech sequences (see section
\ref{sectionlagrangianformulation}) to cochains 
of degree $>2$  without introducing, e.g., gerbes.     
\\ \\
We now study the covariance properties of the subsequences
(\ref{spencer11}) and (\ref{spencer12}), this will give rise to a refined
version of (\ref{firstspencersequence}), called second Spencer sequence in
\cite{pom}. 
\\ \\ 
$\bullet$ First, we study the covariance of (\ref{spencer11}) with respect to
the structure group ${\cal{GL}}_{k+1}$ of the principal bundle 
\begin{equation*}
     \xymatrix{ 
     {\cal{GL}}_{k+1} \ar[r] & 
     \rec{Aut}(M_{k+1}) \ar[r] & 
     \rec{Aut}(M_k) 
     }
\end{equation*}
For a gauge transformation $g_{k+1} \in {\cal{GL}}_{k+1}$ ($k > -1$ otherwise
we get nothing), we have 
\begin{eqnarray}
    D_{\theta}(f_{k+1} \circ g_{k+1}) 
    & = & 
    {g_{k+1}}^* D_{\theta} f_{k+1} + D_{\theta} g_{k+1}  \nonumber \\ 
    & = & 
    \bad{Ad} ({\tilde{g}_{k+1}}^{-1}) D_{\theta} f_{k+1} + 
    \bad{Ad} ({\tilde{g}_{k+1}}^{-1}) \theta^k - \theta^k \nonumber \\
    & = & 
    \bad{Ad} ({\tilde{g}_{k+1}}^{-1}) D_{\theta} f_{k+1} 
    + \tilde{\mu}_k \circ \theta^k  \nonumber \\
    & = & 
    D_{\theta} f_{k+1} + \tilde{\mu}_k \circ {f_{k+1}}^* \theta^k 
    \label{covariancediffmax}
\end{eqnarray}        
where $\tilde{g}_{k+1}$ is the equivariant function corresponding to $g_{k+1}$,
and $\tilde{\mu}_{k}$ is the section of 
$M_{k+1} \times_{\rec{Ad}} \got{gl}_{k+1} \subset 
M_{k+1} \times_{\bad{Ad}} \got{gl}_{k,1}$ 
such that at each point $e_{k+1}$ (see sections \ref{sectionthejetaction} 
and \ref{sectiondeformationspace}): 
$ \bad{Ad} (\tilde{g}_{k+1}) X = X - \tilde{\mu}_k X_{-1} $ (the minus sign is
taken because the gauge transformation $\tilde{g}_{k+1}$ is the particular
deformation $(1 +\tilde{\mu}_k)^{-1} = 1 - \tilde{\mu}_k$ for $k > -1$), 
with $\partial \tilde{\mu}_k = 0$ i.e. $\tilde{\mu}_k{_{|e_{k+1}}} \in
\got{gl}_{k+1} \simeq GL_{k+1}$.  
The covariance law (\ref{covariancediffmax}), which is just the composition of the deformations
$\tilde{D}_{\theta} f_{k+1}$ and $\tilde{\mu}_k$, is rewritten from the
equivariant viewpoint as : 
\begin{equation}
     \tilde{D}_{\theta} (f_{k+1} \circ g_{k+1}) 
     = 
     \tilde{D}_{\theta} f_{k+1} + \tilde{\mu}_k \circ 
     (1 + \tilde{D}_{\theta} f_{k+1})
     \label{firstcovariance} 
\end{equation}   
where all quantities are evaluated at the same $e_{k+1}$, contrary to equation
(\ref{compojetdef}). 
This suggests to define the quotient bundle 
$$ \overline{{\cal{B}}}_k = {\cal{B}}_k / (M_{k+1} \times_{\rec{Ad}} 
   \got{gl}_{k+1})
   \simeq M_{k+1} \times_{\bad{Ad}} (B_k / \got{gl}_{k+1}) 
$$
where $GL^{k+1}$ acts naturally on $B_k / \got{gl}_{k+1}$, and denote by
$\Omega'^1(M,\overline{S}_k)$ its space of sections, which satisfies 
$$ \Omega'^1(M,\overline{S}_k) = \Omega'^1 (M,S_k) / \Gamma (M_{k+1}
\times_{\rec{Ad}} \got{gl}_{k+1}) $$
Note that, at the fiber level we have $B_k / \got{gl}_{k+1} \simeq B_{k-1}
\ltimes (\got{gl}_{k,1} / \got{gl}_{k+1} )$. 
The calculations above then show that the operator 
\begin{eqnarray*}
    \overline{D}_{\theta} 
    &:& 
    \rec{Aut} (M_{k}) \to \Omega'^1 (M,\overline{S}_k)
    \\ 
    && 
    f_{k} \mapsto \overline{D}_{\theta} f_k = 
    \tilde{D}_{\theta} f_{k+1} \circ (1+\tilde{D}_{\theta}f_{k+1})^{-1} \circ
    \theta
    \mod \Gamma (M_{k+1} \times_{\rec{Ad}} \got{gl}_{k+1})         
\end{eqnarray*}
is well defined for any $f_{k+1}$ above $f_k$. \\
Note that the projection map $\Omega'^1(M,S_k) \to \Omega'^1(M,\overline{S}_k)$ is, in
relation with the definition of $\overline{D}_{\theta}$, 
$$
    \tilde{\mu} \to \tilde{\mu} (1 + \tilde{\mu})^{-1} \mod \got{gl}_{k+1} 
$$
as we have (see section \ref{sectiondeformationspace}) $\tilde{\mu} =
[\tilde{D}_{\theta} f_{k+1}]$, 
and the action of $\tilde{\nu}_k \in \Gamma (M_{k+1} \times_{\rec{Ad}}
\got{gl}_{k+1} )$ on $\Omega'^1(M,S_k)$ defining the quotient is : 
$$ \tilde{\mu} \to \tilde{\mu} + \tilde{\nu}_k \circ (1 + \tilde{\mu}) 
$$
which keeps invariant the class of $\mu$. \\  
We can summarise this construction in the exact commutative diagram 
\begin{equation*}
   \xymatrix{
   \rec{id} \ar[r] \ar[d] & 
   {\cal{GL}}_{k+1} \ar[r]^-{D_{\theta} \simeq \bad{Ad}} \ar[d] & 
   \Gamma(M_{k+1} \times_{\rec{Ad}} \got{gl}_{k+1}) \ar[d] \\
   \rec{Aut}(M) \ar[r]^-{j_{k+1}} \ar@{=}[d] & 
   \rec{Aut}(M_{k+1}) \ar[r]^-{D_{\theta}} \ar[d] & 
   \Omega'^1 (M,S_k) \ar[d] \\
   \rec{Aut}(M) \ar[r]^-{j_k} & 
   \rec{Aut}(M_k) \ar[r]^-{\overline{D}_{\theta}} & 
   \Omega'^1 (M,\overline{S}_k)  
   }
\end{equation*}    
where the first line corresponds to the covariance law under ${\cal{GL}}^{k+1}$
and the central row encodes the symmetry we started from. This results in the
sequence of the last line, which is the projected version of (\ref{spencer11}).
By construction, we then end with the exact sequence :   
\begin{equation*}
   \xymatrix{
   \rec{id} \ar[r] &
   \rec{Aut} (M) \ar[r]^-{j_{k}} & 
   \rec{Aut} (M_{k}) \ar[r]^-{\overline{D}_{\theta}} & 
   \Omega'^1(M,\overline{S}_k) 
   }
\end{equation*}
$\bullet$ 
Second, we study the covariance of (\ref{spencer12}) with respect to the
structure group $\Gamma (M_{k+1} \times_{\rec{Ad}} \got{gl}_{k+1})$ of the
principal bundle (which is the third row of the preceding diagram): 
\begin{equation*}
     \xymatrix{
     \Gamma(M_{k+1} \times_{\rec{Ad}} \got{gl}_{k+1}) \ar[r] & 
     \Omega'^1(M,S_{k}) \ar[r] & 
     \Omega'^1(M,\overline{S}_k) 
     }
\end{equation*} 
and more generally under the group $\Gamma (M_{k+1} \times_{\bad{Ad}}
\got{gl}_{k,1}) \subset 
\Omega'^1(M,S_{k}) \simeq \Gamma ({\cal{B}}_k) $. 
Inspired by the preceding point, the action of a maximal degree deformation 
$\nu_{k} \in  \Gamma(M_{k+1} \times_{\bad{Ad}}
\got{gl}_{k,1}) $ for $k>-1$ is given by :
$$ \tilde{\mu} \to \tilde{\mu} + \tilde{\nu}_k  \circ (1 + \tilde{\mu} )$$ 
that is  
$$ \mu \to \mu + \nu_k + i_{\mu} \nu_k $$
in form language. 
Next, a direct calculation gives (this is another version of the cocycle law 
(\ref{defcocycle})) 
\begin{eqnarray}
     D_{\theta} (\mu + \nu_{k} + i_{\mu} \nu_k) 
     &=& 
     D_{\theta} \mu + [ \theta + \mu  , \tilde{\nu}_k \circ (\theta + \mu )] \mod \got{h}_{k-1} 
     \nonumber \\
     & = &
     D_{\theta} \mu + \partial \tilde{\nu}_k \circ (\theta + \mu) 
     \label{covariancedefmax}
\end{eqnarray}
From the
equivariant viewpoint, the covariance law (\ref{covariancedefmax}) reads
(compare with equation (\ref{firstcovariance})) 
\begin{eqnarray}
    \tilde{D}_{\theta} (\mu. \nu_k) &=&
    \tilde{D}_{\theta} \mu + 
    \partial \tilde{\nu}_k  \circ (1 + \tilde{\mu})
    \label{secondcovariance} \\
    &=& 
    \tilde{D}_{\theta} \mu + 
    \partial \tilde{\nu}_k  
    \circ (1 + \tilde{\mu}_{-1})
    \nonumber
\end{eqnarray}
This suggests to define the quotient bundle     
\begin{equation*}
     \Lambda^2 (M,S_{k-1}) / (M_{k} \times_{\bad{Ad}} \partial \got{gl}_{k,1} )
     \simeq 
     M_{k} \times_{\bad{Ad}} (\got{gl}_{-1,2} \oplus \cdots \oplus 
     \got{gl}_{k-1,2})/ \partial \got{gl}_{k,1} 
\end{equation*}
whose space of sections, denoted $\Omega^2 (M,\overline{S}_{k-1})$, satisfies : 
\begin{equation*}
     \Omega^2 (M,\overline{S}_{k-1}) = \Omega^2 (M,S_{k-1}) / 
     \Gamma (M_k \times_{\bad{Ad}} \partial \got{gl}_{k,1}) 
\end{equation*}
The preceding calculations then proves that if $\nu_{k}$ is a deformation in the
structure group $\Gamma(M_{k+1} \times_{\rec{Ad}} \got{gl}_{k+1})$, i.e.
$\partial \tilde{\nu}_k = 0$, then $D_{\theta} \mu$ is left invariant under its
action, and that the operator 
\begin{eqnarray*}
    \overline{D}_{\theta} 
    &:& 
    \Omega'^1 (M,\overline{S}_k) \to 
    \Omega^2 (M,\overline{S}_{k-1})  \\
    & &
    \overline{\mu} \mapsto 
    \overline{D}_{\theta} \overline{\mu} = 
    \tilde{D}_{\theta} \mu \circ (1 + \tilde{\mu})^{-1} \circ \theta
    \mod \Gamma (M_k \times_{\bad{Ad}} \partial \got{gl}_{k,1})
\end{eqnarray*}
is well defined for any $\mu$ above $\overline{\mu} \in \Omega'^1
(M,\overline{S}_k)$. \\ \\ 
The construction is summarised in the commutative diagram  
\begin{equation*}
    \xymatrix{ 
    {\cal{GL}}_{k+1} \ar[r]^-{D_{\theta} \simeq \bad{Ad}}_-{\simeq} 
                     \ar[d] & 
    \Gamma( M_{k+1} \times_{\rec{Ad}} \got{gl}_{k+1}) \ar[r]^-{D_{\theta} \simeq
    \partial} 
      \ar[d] & 
    \Gamma (M_{k} \times_{\bad{Ad}} \partial \got{gl}_{k,1} ) \ar[d]  
    \\
    \rec{Aut}(M_{k+1}) \ar[r]^-{D_{\theta}} \ar[d] & 
    \Omega'^1(M,S_k) \ar[r]^-{D_{\theta}} \ar[d] & 
    \Omega^2 (M,S_{k-1}) \ar[d] 
    \\ 
    \rec{Aut}(M_k) \ar[r]^-{\overline{D}_{\theta}} & 
    \Omega'^1(M,\overline{S}_k) \ar[r]^-{\overline{D}_{\theta}} & 
    \Omega^2(M,\overline{S}_{k-1}) 
    }
\end{equation*}    
where the first line corresponds to the covariance under maximal degree
deformations, and the second row encodes the symmetry we started from. 
This gives the projected version of
(\ref{spencer12}), that is the sequence : 
\begin{equation*}
    \xymatrix{
    \rec{Aut}(M_{k}) \ar[r]^-{\overline{D}_{\theta}} & 
    \Omega'^1(M,\overline{S}_k) \ar[r]^-{\overline{D}_{\theta}} &
    \Omega^2(M,\overline{S}_{k-1}) 
    }
\end{equation*}    
\\       
$\bullet$ 
Putting things altogether, we obtain thus the non linear second Spencer sequence : 
\begin{equation*}
    \xymatrix{
    \rec{id} \ar[r] &
    \rec{Aut}(M) \ar[r]^-{j_{k}} & 
    \rec{Aut} (M_k) \ar[r]^-{\overline{D}_{\theta}} & 
    \Omega'^1(M,\overline{S}_k) \ar[r]^-{\overline{D}_{\theta}} & 
    \Omega^2(M,\overline{S}_{k-1}) 
     }
\end{equation*}
This is the projected form of (\ref{firstspencersequence}).
\\ \\
$\bullet$ Finally, note that the linearised version of the first Spencer
sequence (\ref{firstspencersequence}) is (we still denote $j_{k+1}$ the
linearised version)  
\begin{equation*}
     \xymatrix{
     0 \ar[r] & 
     \rec{aut}(M) \ar[r]^-{j_{k+1}} & 
     \rec{aut}(M_{k+1}) \ar[r]^-{d_{\theta}} & 
     \Omega^1 (M,S_k) \ar[r]^-{d_{\theta}} & 
     \Omega^2 (M,S_{k-1}) 
     }  
\end{equation*}
$\rec{aut}(M) \simeq \Gamma(TM)$ is the Lie algebra of $\rec{Aut}(M)$
i.e. the vector fields on $M$ which satisfies 
$$ \rec{aut}(M) \simeq \Omega^0 (M,S_{-1}) $$ 
$\rec{aut}(M_{k+1})$ is the Lie algebra of
$\rec{Aut}(M_{k+1})$ i.e. the right invariant vector fields on $M_{k+1}$ which
satisfies : 
$$ \rec{aut} (M_{k+1}) \simeq \Omega^0 (M,S_{k+1}) $$ 
So, this linearised sequence contains the beginning of the linear sequence 
(\ref{linearspencersequence1}). Putting these together, we obtain the linear
Spencer sequence : 
\begin{eqnarray*}    
    &\xymatrix{ 
    0 \ar[r] &
    \Omega^0(M,S_{-1}) \ar[r]^-{j_{k+1}} &
    \Omega^0 (M,S_{k+1}) \ar[r]^-{d_{\theta}} &
    \Omega^1 (M,S_{k}) \ar[r]^-{d_{\theta}} &    
    } \,\,\, \,\,\, \,\,\, \,\,\, \qquad  \qquad  \qquad  &
    \\     
    & \,\,\, \,\,\, \,\,\, \,\,\,  \,\,\, \,\,\, \,\,\, \,\,\, 
    \qquad \qquad \qquad  \qquad  \qquad  \qquad \qquad 
    \xymatrix{
    \cdots \ar[r] & 
    \Omega^n (M,S_{k+1-n}) \ar[r]^-{d_{\theta}} &
    0 
    }&
\end{eqnarray*}
This sequence is locally exact \cite{pom}.  

\subsubsection{Lagrangian and \u{C}ech formulations}
\label{sectionlagrangianformulation}

$\bullet$ 
On the differentiable $n$-manifold $M$, we consider the lagrangian 
\begin{equation}
     {\cal{L}} (\beta , \mu) = \rec{tr} \,\, \beta \wedge D_{\theta} \mu
\end{equation}
for $\mu \in \Omega'^1(M,S_k)$, and 
$\beta \in \Omega^{n-2} (M , S_{k-1}^*)$. $\rec{tr}$
is the coupling between $\got{gl}_{-1} \oplus \cdots \oplus \got{gl}_{k-1}$ 
and its dual, and $S_{k-1}^*$ is the dual vector bundle of $S_{k-1}$. 
This lagrangian is analogue to the $bc$ models of 2D CFT and to the 
$BF$ models of gauge theory \cite{cotta}. 
\\ \\
The lagrangian $\cal L$ has $\rec{Aut}(M_{k+1})$ symmetry : 
\begin{equation} 
    \mu \to f_{k+1}{^*} \mu + D_{\theta} f_{k+1} , \,\,\, 
    \beta \to f_{k+1}{^*} \beta \,\,\,  
    \Longrightarrow \,\,\, 
    {f_{k+1}}^* {\cal{L}} = {\cal{L}}
    \label{symmetry1}
\end{equation}
since $D_{\theta} \mu \to {f_{k+1}}^* D_{\theta} \mu$ under the action of
$f_{k+1} \in \rec{Aut}(M_{k+1})$. 
The equations of motions are : 
\begin{equation}
     D_{\theta} \mu = 0 , \,\,\, d_{\theta + \mu }^* \beta = 0 
     \label{eqmotion}
\end{equation}
Here the dual $d_{\omega}^*$ of $d_{\omega}$, $\omega = \theta + \mu$, is
defined by 
$$ 
     d \, \rec{tr} \,\, \beta \wedge \alpha = 
     \rec{tr} \,\, d_{\omega}^* \beta \wedge \alpha + (-1)^{n-2} 
     \rec{tr} \,\, \beta \wedge d_{\omega} \alpha 
$$
for all $\alpha \in \Omega^1 (M,S_{k})$.      
\\ \\      
We see that (\ref{eqmotion}) corresponds to the fact that the lagrangian $\cal L$
computes non linear Spencer cocycles and (\ref{symmetry1}) corresponds to the
covariance property of the non linear Spencer sequence under
$\rec{Aut}(M_{k+1})$. Both combined proves
that $\cal L$ is indeed computing non linear Spencer cohomology at the
$\Omega'^1(M,S_k)$ level. 
Of course, one can similarly define a lagrangian model relative to the
linear Spencer sequence. 
\\ \\
$\bullet$
Now, we shall end by a calculation emphasizing the analogy between $\cal{L}$
and $BF$ gauge theory models \cite{cotta}, that is between $k$-frames and gauge theory. \\
Either from the lagrangian, or from the Spencer sequence point of vue, the 
equation of motion for the deformation 
$$ D_{\theta} \mu  = 0 $$
is locally solved by 
\begin{equation}
    \mu = D_{\theta} f_{k+1,i} 
    \label{solutionBF}
\end{equation}
for $f_{k+1,i} \in \rec{Aut}(U_{i,k+1})$ above a open subset $U_i \subset M$.
The $U_i$'s are chosen as in section \ref{synthesisdeformation}. 
As $\mu$ is globally defined, equation (\ref{solutionBF}) implies that, above 
$U_{ij} = U_i \cap U_j$, we have 
$ 
    D_{\theta} f_{k+1,i} = D_{\theta} f_{k+1,j} 
$, so the element $f_{k+1,ij} = f_{k+1,i} \circ {f_{k+1,j}}^{-1} \in
\rec{Aut}(U_{ij,k+1})$ satisfies, thanks to the cocycle property of $D_{\theta}$ : 
\begin{equation*}
     D_{\theta} f_{k+1, i} = 
     D_{\theta} (f_{k+1,ij} \circ f_{k+1,j} ) = 
     {f_{k+1,j}}^* D_{\theta} f_{k+1,ij} + D_{\theta}
     f_{k+1,j} 
     \,\,\, 
     \Longrightarrow 
     \,\,\, 
     D_{\theta} f_{k+1,ij} = 0  
\end{equation*}             
so we have $ f_{k+1,ij} = j_{k+1} (f_{-1,ij}) $ (exactness of
(\ref{firstspencersequence})) where $f_{-1,ij} = f_{ij}$ is a
diffeomorphism of $U_{ij}$. 
Next, we also have 
$$
     f_{k+1,ij} \circ f_{k+1,jk} \circ f_{k+1,ki} = \rec{id} \,\,\, , \,\,\,
     {\rm above} \,\,\, U_{ijk} = U_i \cap U_j \cap U_k 
$$
so, as $j_{k+1}$ is a morphism, 
$$
     j_{k+1} (f_{ij} \circ f_{jk} \circ f_{ki}) = \rec{id} \,\,\, , \,\,\, 
     {\rm above} \,\,\, U_{ijk} 
$$
Now, as $j_{k+1}$ is injective (exactness of (\ref{firstspencersequence}) again), 
this last equality is equivalent to 
$$
     f_{ij} \circ f_{jk} \circ f_{ki} = \rec{id} \,\,\, , \,\,\, 
     {\rm on} \,\,\, U_{ijk}                
$$
Consequently we have associated to $\mu$ a \u{C}ech 1-cocycle $(f_{ij})$ with
values in the diffeomorphisms of $M$.
\\ \\
Note that the same type of calculation proves that
$f_{k+1,i}$ is defined up to the transformation 
$$ f_{k+1,i} \to j_{k+1}(f'_{-1,i}) \circ f_{k+1,i} \,\,\, , \,\,\, {\rm{for}} 
\,\,\, f'_{-1,i} \in \rec{Aut}(U_i) $$
because of the cocycle property : 
$$ D_{\theta} (j_{k+1} (f'_{-1,i}) \circ f_{k+1,i}) = 
   {f_{k+1,i}}^* D_{\theta} (j_{k+1} (f'_{-1,i})) + D_{\theta} f_{k+1,i}
   = 
   D_{\theta} f_{k+1,i}
$$
Under such a transformation, the \u{C}ech cochains transform as 
\begin{eqnarray*}
     f_{k+1,ij} &\to& j_{k+1} (f'_{-1,i}) \circ f_{k+1,ij} \circ j_{k+1} 
     ({f'_{-1,j}}^{-1}) \\
     f_{-1,ij} &\to& f'_{-1,i} \circ f_{-1,ij} \circ {f'_{-1,j}}^{-1} 
\end{eqnarray*}     
These covariance properties are the \u{C}ech version of the covariance under
$\rec{Aut}(M)$ of the non linear Spencer sequence, or alternatively of the space
(\ref{productspace}). 
\\ \\
$\bullet$
All these facts suggest that the (differential) cohomology of the non linear Spencer 
sequence is related to the (combinatorial and non abelian) 
cohomology of diffeomorphisms \u{C}ech type
sequences. Recall what are the \u{C}ech cochains for the
diffeomorphisms. 0-cochains are 
$ (f_i) \in C^0 (\rec{Aut}(M)) $  
where $f_i$ is a diffeomorphism of $U_i$, 1-cochains are 
$ (f_{ij}) \in C^1 (\rec{Aut}(M)) $ 
where $f_{ij}$ is a diffeomorphism of $U_{ij}$ with $f_{ji} = {f_{ij}}^{-1}$, 
and 2-cochains are 
$ (f_{ijk}) \in C^2(\rec{Aut}(M)) $
where $f_{ijk}$ is a diffeomorphism of $U_{ijk}$. The \u{C}ech differential
$\delta $ is defined as usual, respectively on 0-cochains and 1-cochains by : 
\begin{eqnarray*}
      (\delta f)_{ij} &=& f_i \circ {f_{j}}^{-1} \\    
      (\delta f)_{ijk} &=& f_{ij} \circ f_{jk} \circ f_{ki} 
\end{eqnarray*}
With this, using holonomy/homotopy type arguments, we expect that the  
cohomology of the \u{C}ech sequence (the second arrow being the 
restriction map)
\begin{equation*}
    \xymatrix{
    \rec{id} \ar[r] &
    \rec{Aut}(M) \ar[r] & 
    C^0(\rec{Aut}(M)) \ar[r]^-{\delta} &
    C^1(\rec{Aut}(M)) \ar[r]^-{\delta} &
    C^2(\rec{Aut}(M))  
    }
\end{equation*}    
is isomorphic to the Spencer non linear cohomology. 
\\ \\ 
$\bullet$ 
Of course, the interest in the lagrangian $\cal{L}$ is as limited as those of
$BF$ type in gauge theory : it only encodes topological information on the space
$M$ equiped with a background differential structure. Nevertheless, we formally
expect, as in \cite{cotta} for gauge theory, that the quantum theory corresponding to $\cal{L}$ is encoded in some
sort of non abelian intersection theory between $1$-cycles (sources of the $\mu$
field) and $(n-2)$-cycles (sources of the $\beta$ field) in $M$, the cycles
being here understood in the sense of some non abelian singular homology.    
\\ \\
$\bullet$ 
The theory of linear frames, in all the aspects described here, as well as
another ones like e.g. flag structures \cite{cap}, can be modified (reduction of
frame bundles) or extended (definition of graded type frames) to embody all
kind of gravitational type structures. The gravitationnal field is then a Cartan
connection, \cite{cap,koba,newman,sharpe}, which can be thought as a $\mu$ field,
or the inverse of some $k$-frame, with $k=2$ for Riemannian gravity, $k=3$ for
conformal \cite{newman} or projective gravity, $k=\infty$ for Kodaira-Spencer gravity
\cite{serge}. 
\\ \\ \\ 
{\bf{Acknowledgments}}
\\ \\
The author thanks S. Lazzarini for discussions and reading of the manuscript.

\end{document}